\documentclass[twocolumn]{aastex61}
\usepackage{comment}
\usepackage{bm}
\usepackage{amsmath}

\submitjournal{ApJ}

\shorttitle{Incidence of AGNs  in $z<2.5$ Mergers}
\shortauthors{Silva et al.}
\begin{document}
\title{Galaxy mergers up to $z<2.5$ II:  agn incidence in merging galaxies at separations of 3-15 kpc}
\correspondingauthor{Silva et al.}
\email{anggitawins@gmail.com}

\author{Andrea Silva}
\affil{National Astronomical Observatory of Japan, National Institutes of Natural Sciences, 2-21-1 Osawa, Mitaka, Tokyo 181-8588, Japan}
\author{Danilo Marchesini}
\affil{Department of Physics and Astronomy, Tufts University, Medford, MA 02155, USA}
\affil{International Research Fellow of Japan Society for the Promotion of Science (Invitational Fellowships for Research in Japan — Short-term}
\author{John D. Silverman}
\affil{Kavli Institute for the Physics and Mathematics of the Universe (WPI), The University of Tokyo Institutes for Advanced Study, The University of Tokyo, Kashiwa, Chiba 277-8583, Japan}
\author{Nicholas Martis}
\affil{Department of Physics and Astronomy, Tufts University, Medford, MA 02155, USA}
\author{Daisuke Iono}
\affil{National Astronomical Observatory of Japan, National Institutes of Natural Sciences, 2-21-1 Osawa, Mitaka, Tokyo 181-8588, Japan}
\author{Daniel Espada}
\affil{SKA Organization, Lower Withington, Macclesfield, Cheshire SK11 9DL, UK}
\author{Rosalind Skelton}
\affil{South African Astronomical Observatory, PO Box 9, Observatory, Cape Town 7935, South Africa}

%%%%%  ABSTRACT %%%%%%%
\begin{abstract}
We present a study of the incidence of active galactic nucleus (AGN) in a sample of  major merging systems at 0.3$<$$z$$<$2.5. Galaxies in this merger sample have projected separations between 3 to 15 kpc and are selected from the CANDELS/3D-HST catalogs using a peak-finding algorithm. 
AGNs in mergers and non-mergers are identified on the basis of their X-ray emission, optical lines, mid-infrared colors, and radio emission. 
Among galaxies with adequate measurements to find potential AGNs, we 
 find a  similar fraction of AGNs in mergers (16.4$\pm ^{5.0}_{3.1}$\%) compared to the fraction found in non-merging galaxies (15.4$\pm$0.6\%).  In mergers, this fraction is obtained by assuming that, in unresolved observations, only one of the merging galaxies is the AGN source. The similarity between the fractions is possibly due to the higher availability of cold gas at high redshifts,  where the excess of nuclear activity as a result of merging is less important than at lower redshifts. 
Star-forming galaxies have a higher incidence of AGNs than quiescent galaxies. In particular, starbursts in mergers are the most common sites of 
AGN activity since they present higher AGN fractions and black hole accretion rates.
We find no clear correlation between the black hole accretion rate and the galaxy properties (i.e., 
star-formation rate, stellar mass) in mergers and non-mergers. 
However,  mergers seem to have a higher correlation with star formation than non-mergers, which possibly indicates that  the merging process is starting to influence the star formation and AGN activity even at this pre-coalescence stage.

\end{abstract}

%%%%%   INTRODUCTION  %%%%%%%
\section{Introduction} \label{sec_intro}

Almost all massive galaxies harbor  a supermassive black hole (SMBH; 10$^5$$<$M$_{\rm BH}$/M$_{\odot}$$<$10$^{10}$) in their centers  \citep{kormendy1995, kormendy2013}, which grow via heavy accretion events. When material falls into the accretion disk surrounding the SMBH, huge amounts of energy can be released as radiation over a wide range of wavelengths, making the galaxy to shine as an active galactic nucleus (AGN).  

 In order to accrete onto a SMBH, matter in a galaxy needs to  lose 99\% of its angular momentum to move from kpc scales to the inner regions of the galaxy center to reach  the vicinity of the accretion disk \citep{shlosman1990}. 
One of the main mechanisms proposed to explain the ignition of nuclear activity is the merging of two or more galaxies with similar mass also known as major mergers  \citep{sanders1988, silk1998, springel2005}. 
Numerical simulations of mergers \citep[e.g.][]{hernquist1989, barnes1996} suggest that during the interaction of  two galaxies, the gaseous content in them feels a strong gravitational torque from tidally-induced stellar bars. The gas loses angular momentum and flows inwards the nuclei of the galaxies activating  rapid accretion onto their respective SMBHs.  In the final fusion of the merging galaxies, a second phase of inflow occurs, as gas is driven further inwards to the final nucleus of the resultant galaxy.

Although a number of  merger simulations  show that merging is a very efficient way to trigger black hole growth in a short timescale \citep[e.g.][]{dimatteo2005, hopkins2008}, observationally, the link between mergers and the  triggering of AGN activity is not clear.  
This connection has been favored by the findings of a high fraction of AGN hosts with morphologies suggesting recent merger activities \citep{koss2010, koss2012,  bessiere2012, cotini2013, goulding2018, ellison2019} and an excess on the AGN fraction in merging galaxies and galaxy pairs compared to control samples \citep{alonso2007, woods2007, ellison2011, silverman2011, lackner2014, satyapal2014, weston2017, ellison2019}, suggesting that mergers are capable of triggering nuclear activity. . 
However,   there is also a significant number of works that  have severely questioned this connection \citep{gabor2009, cisternas2011, kocevski2012, villforth2014, rosario2015, sabater2015, mechtley2016, villforth2017, marian2019}. 
It is suggested that the majority of the AGNs, especially the  low- to mid-luminosity  AGNs, are triggered by processes other than mergers  (such as secular processes), while  major mergers trigger only the most luminous AGNs \citep{ urrutia2008, treister2012, menci2014, glikman2015, fan2016,  weigel2018}, although recent observations of 
\citet{villforth2017} and \citet{hewlett2017} question such a relation. 

In addition of triggering AGN activity, major mergers are expected to enhance the star formation in the system \citep{kartaltepe2012, yuan2012, hung2013, lanz2013, patton2013}.  This picture  is supported by observations of local starbursts since the majority of them show disturbed morphologies as a result of recent merger activity \citep{armus1987, sanders1996, kartaltepe2010,
ellison2013,  espada2018}. 
 However,  as predicted by recent simulations \citep{perret2014, scudder2015, fensch2017}, high-redshift  galaxy mergers  have been observed to be inefficient at enhancing the star formation activity in the system and producing starbursts \citep{xu2012, kaviraj2013, kaviraj2015, lofthouse2017, silva2018, pearson2019, wilson2019}, therefore it is also possible that high-$z$ mergers are inefficient at triggering AGN activity.  
However,  studies  of  the effects of mergers at high redshifts on igniting starburst and AGN activities are scarce due to the difficulty of finding samples of high-$z$ mergers.  Some of the methods  to find them include the selection of galaxy pairs (at close projected separations and redshift), which potentially are mergers before coalescence \citep[e.g.][]{robaina2010, williams2011, tasca2014, lackner2014};  the selection of  galaxies 
 with morphologies suggesting recent or post-merger activity \citep[e.g.][]{bridge2010, cisternas2011, kartaltepe2015}; and the use of deep learning techniques which perform visual-like classification of merging galaxies  \citep[e.g.][]{pearson2019}. 
 
Another important topic to explore observationally is the existence (or lack) of the correlation 
between the black hole accretion rate (BHAR) and the star formation rate (SFR) in mergers, to analyze the level of star formation and black hole accretion they can induce and compare with the level produced by non-merging galaxies.
This topic arised from the observed correlation between the mass of the stellar bulge in a galaxy (and also the velocity dispersion) and the mass of its central supermassive black hole \citep{magorrian1998, ferrarese2000, gebhardt2000, marconi2004} (with a local correlation value of M$_{\rm BH}$/M$_{\rm bulge}$$\sim$1.5$\times10^{-3}$, \citealt{mclure2002}), suggesting a parallel evolution between the black hole and the stellar mass of the host galaxy.  
For this co-evolution to take place, the BHAR should be proportional to the SFR in the galaxy, especially the star formation rate that builds up the bulge.  Both the black hole accretion and the star formation  in a galaxy need a common fuel of cold gas supply to arise. However, the spatial scales at which these processes take place  are vastly different (many kpc and sub-pc scales for star formation and black hole accretion, respectively).  
Since  galaxy mergers have been proposed as the main mechanism that trigger the loss of angular momentum of the cold gas from kpc to pc scales,
it is expected that the resultant galaxies will have a SMBH and the stellar content growing in a parallel way. In contrast, galaxies growing through secular processes should have unrelated black hole and stellar growth. 
Observationally, the link between the star formation rate and the black hole accretion rate is debated, with some works indicating that  at low AGN luminosities (L$_{\rm bol}$$<$10$^{45}$ erg s$^{-1}$), 
these rates are uncorrelated \citep{shao2010, lutz2010, rovilos2012, rosario2012},
while in  high-luminosity AGNs (which are possibly triggered by mergers) a significant correlation emerges \citep{lutz2008, serjeant2010}, although some works also find weak or inverted connections \citep{page2012, harrison2012}.
The feedback model of \citet{silk2013}, which couples star formation  and black hole accretion via outflow-induced pressure-enhanced star formation, predicts an   average BHAR/SFR$\sim$10$^{-3}$.
On the contrary, \citet{volonteri2015} use hydrodynamical simulations of galaxy mergers at different merger stages and find no strong correlation between the BHAR and the SFR.

In this paper, we  investigate the role of mergers (major mergers with mass ratio $>$1:4) in triggering black hole growth at 0.3$<$$z$$<$2.5.  
 We study AGN activity in the  sample of major mergers obtained in  \citet{silva2018}, which is the first paper of the series.  In \citet{silva2018},  we investigated  the influence of galaxy mergers on star formation by studying a  sample of  merging galaxies with nuclei projected separations between 3 to 15 kpc and in the redshift range of 0.3$<$$z$$<$2.5.   We compared the star formation activity between mergers and non-mergers and  found no significant difference between them suggesting that high redshift mergers at this stage are possibly less efficient than their local counterparts at enhancing the star formation activity in galaxies.
  Here, we identify AGNs in  this sample of mergers by analyzing  their X-ray and radio luminosity,  optical line emissions,  and mid-infrared colors, and compare their AGN activity with non-merging galaxies. 
  Section \S 2 presents the sample and the multi-wavelength data used to identify AGNs, while Section \S 3 details the adopted AGN selection criteria. Section \S 4 presents the results, which are then discussed in Section \S 5. Finally, Section \S 6 summarizes the paper and its conclusions. 
Throughout this paper, we adopt a cosmology with $H_{0}$=70 km s$^{-1}$ Mpc$^{-1}$, $\Omega_{\Lambda}$=0.7, and $\Omega_{m}$=0.3.  Magnitudes are in the AB system.

%%%%%   DATA AND SAMPLE  %%%%%   
\section{Data} \label{sec_data} \label{sec_data_sel}

 \subsection{Merger Sample} \label{sec_merg_sample}

Full details on the selection of galaxy mergers are described  in \citet{silva2018}. Here we only present the key aspects:

Galaxy pairs were selected by applying the technique described in \citet{lackner2014} to near-IR  H160 postage stamps.  
These postage stamps were centered on 5717  galaxies with log(M$_{\star}$/M$_{\odot}$)$\ge$10,  $m_{\rm AB}$$<$24.5, and  0.3$<$$z$$<$2.5, which were  obtained from  the  3D-HST catalogs \citep{skelton2014, momcheva2016}. 
We selected galaxy pairs with nuclei separations between 3 to 15 kpc and with mass and  flux ratios $\ge$1:4 to later select potential major mergers.  We found that 28\% of the selected galaxy pairs were unresolved in the 3D-HST catalogs (thus the total number of sources in the sample increased to 5907\footnote{This number includes galaxies with masses $\log(\rm M_{\star}/M_{\odot})<$10.}). We extracted deblended photometry for these galaxies and derived photometric redshifts, colors, and stellar population properties. 

We used the redshift values of the galaxies in pairs, provided by the 3D-HST catalogs  and the deblended photometry (photometric, grism, and spectroscopic redshifts) to separate potential mergers  of  line of sight contaminants. We defined mergers as galaxies that are consistent with being at the same redshift within a 3$\sigma$ uncertainty (when using either photometric or grism redshifts), or if the redshift differ by less than 0.001 (if both have spectroscopic redshit).
We found 130 major merging systems  in which the merging galaxies have masses in the range 8.3$\le$$\log(\rm M_{\star}/M_{\odot})$$\le$11.5 (primary sample).  Since star-forming clumps in galaxies usually have stellar masses 8$\le$$\log(\rm M_{\star}/M_{\odot})$$\le$10 \citep{guo2012}, we constructed a more restrictive sample of major mergers (high-mass sample) in which both merging galaxies have stellar masses log(M$_{\star}$/M$_{\odot}$)$\ge$10\footnote{The completeness in stellar mass at log(M$_{\star}$/M$_{\odot}$)$\ge$10 is above 90\% at $z$=2.5.}. This high-mass sample contains 64 merging systems (128 galaxies).  The sample of non-merging galaxies (i.e., galaxies not selected as mergers) with masses log(M$_{\star}$/M$_{\odot}$)$\ge$10 consists of 5718 galaxies.  
In this work, we focus the study on the  incidence of AGNs in mergers using the high-mass sample. 

We used the rest-frame U-V and V-J colors (or UVJ colors) to separate merging galaxies into quiescent, unobscured star-forming, and dusty star-forming  by using the criteria defined in \citet{whitaker2015} and \citet{martis2016}.   As found in \citet{silva2018},  for the 128 galaxies in the merger sample, 35.9$\pm$5.3\%, 21.9$\pm$3.4\%, and 42.2$\pm$5.7\% are quiescent, unobscured, and dusty star-forming galaxies, respectively. For non-mergers the fractions are  30.5$\pm$0.7\%, 28.8$\pm$0.7\%, and 40.7$\pm$0.8\%, respectively. 

 We separated mergers into  wet (both galaxies are star-forming, i.e., dusty or unobscured), mixed (one quiescent and one star-forming galaxy), and dry (both galaxies are quiescent) and  found that 53.1$\pm$6.4\%, 21.9$\pm ^{4.9} _{3.4}$\%, and 25.0$\pm ^{5.2} _{3.7}$\% of the major mergers correspond to wet, mixed, and dry mergers, respectively. 

 Star formation rates were obtained from a combination of rest-frame UV emission and mid-IR photometry obtained from {\it Spitzer}/MIPS imaging following \citet{whitaker2012}. When MIPS 24 $\mu$m data were not available or for detections with S/N$<$3, the SFRs were obtained from the modeling of the spectral energy distributions (SEDs).

We found no difference in the star formation activity between merging and non-merging galaxies and only 12$^{+5}_{-3}$\% of the star-forming galaxies in mergers (6.5$\pm$0.4\% in non-mergers) are a  starburst galaxy, i.e., a galaxy that lies 0.5 dex above the main sequence fit of \citet{whitaker2014}.  All of these starbursts are in wet mergers and are dusty star-forming galaxies. 
Note that some of the starburst galaxies in the sample of  non-mergers are possibly mergers  in the coalescence phase.

\subsection{X-ray Data} \label{xray_data}

We use publicly available X-ray catalogs to cross-match the merger sample and non-merging galaxies with X-ray sources. 
We use
the AEGIS-X Survey (The {\it Chandra} Deep Survey of the Extended Groth Strip,
 \citealt{laird2009}), the {\it Chandra} COSMOS legacy Survey \citep{civano2016, marchesi2016}, 
 the  CDF-N (2 Ms {\it Chandra} Deep Field-North Survey, \citealt{xue2016}) 
 the CDF-S survey (7 Ms {\it Chandra} Deep Field-South Survey, \citealt{luo2017}), 
 the {\it Chandra} counterparts of CANDELS GOODS-S sources \citep{cappelluti2016}, 
  the X-UDS survey (The {\it Chandra} Legacy Survey of the UKIDSS Ultra Deep Survey Field, \citealt{kocevski2018}),
  the  {\it Chandra} COSMOS Survey \citep{elvis2009}, 
  and the ZFOURGE catalogue of AGN candidates  \citep{cowley2016}.

We cross-match merging and non-merging galaxies with the closest X-ray source found in the catalogs. 
These sources have fluxes above the flux limit of the respective catalog from where they were extracted. 
The separation used for matching is equal to the Point Spread Function (PSF) of the observations performed to obtain the catalogs (in the range 1\farcs5$-$5\farcs0).
 When the catalogs indicate that the X-ray sources have an optical counterpart, we use a matching separation of  1\arcsec with respect to the optical position.   

Of the 64 merging systems, 14 (21.9$\pm^{7.5}_{4.4}$\%) are matched with an X-ray source. These mergers have galaxies with projected separations between 4.1$-$14.3 kpc (0\farcs5-1\farcs9), therefore we are not able to  distinguish which of the merging galaxies is the source of X-ray emission. 
 The X-ray matched sources have fluxes in the 0.5$-$7 keV band of $f_{x}$=3.8$\times 10^{-17}-$ 2.4 $\times 10^{-14}$  erg s$^{-1}$ cm$^{-2}$ and  are at redshifts 0.68$<$$z$$<$2.13, corresponding to X-ray luminosities of  $L_{x}$=4.2$\times$10$^{41} -$ 3.9$\times$10$^{44}$ erg s$^{-1}$.
In the case of non-merging galaxies, 555 (9.7$\pm$0.4\%) are identified with an X-ray source.   The sources have fluxes in the range  $f_{x}$=1.2$\times 10^{-17} -$ 1.3$\times$10$^{-13}$  erg s$^{-1}$ cm$^{-2}$ and redshifts  0.31$<$$z$$<$2.49, corresponding to X-ray luminosities $L_{x}$=1.4$\times$10$^{40}-$ 5.7$\times$10$^{44}$ erg s$^{-1}$.

\subsection{Optical Line Data} \label{optical_data}

The 3D-HST catalogs provide optical line measurements such as  [OIII]$\lambda$5007 at 1.4$<$$z$$<$2.2 and  H$\beta$  at 1.5$<$$z$$<$2.3 which can be used to find potential AGNs. We select only galaxies in which the [OIII] and H$\beta$ fluxes  are available and have a S/N$\ge$3, as we will use the mass-excitation diagram of \citet{juneau2014} to identify AGNs in \S  \ref{sec_opt_agn}. 
 We find that 6 individual merging galaxies (in 5 systems) have both [OIII] and H$\beta$ measurements. These galaxies are at 1.68$<$$z$$<$2.1 and have projected separations between 4.8-13.8 kpc. They have  [OIII] fluxes in the range $f_{\rm [OIII]}$=3.8$\times$10$^{-17}-$5.6$\times$10$^{-16}$ erg s$^{-1}$ cm$^{-2}$  which correspond to luminosities $L_{\rm [OIII]}$=2.4$\times$10$^{42}-$5.5$\times$10$^{43}$ erg s$^{-1}$, and H$\beta$ fluxes  $f_{\rm H\beta}=$2.5$\times$10$^{-17}-$9.0$\times$10$^{-17}$ erg s$^{-1}$ cm$^{-2}$ and luminosities $L_{\rm H\beta}$=1.7$\times$10$^{42}-$8.9$\times$10$^{42}$ erg s$^{-1}$. 
For non-mergers, 228 galaxies have both [OIII] and H$\beta$ measurements and are at 1.4$<$$z$$<$2.3. The galaxies have [OIII] fluxes in the range $f_{\rm [OIII]}$=2.1$\times$10$^{-17}-$ 5.6$\times$10$^{-15}$ erg s$^{-1}$ cm$^{-2}$ which correspond to luminosities $L_{\rm [OIII]}$= 1.1$\times$10$^{42}-$ 3.7$\times$10$^{44}$ erg s$^{-1}$, and have H$\beta$ fluxes in the range $f_{\rm H\beta}$=1.1$\times$10$^{-17}- $3.0$\times$10$^{-15}$ erg s$^{-1}$ cm$^{-2}$ corresponding to a range in  luminosities $L_{\rm H\beta}$= 7.7$\times$10$^{41}-$ 2.4$\times$10$^{44}$ erg s$^{-1}$.

\subsection{Mid-Infrared Data}

{\it Spitzer}/IRAC  3.6, 4.5, 5.8, 8.0 $\mu$m fluxes for our sample of mergers and non-mergers are provided by the 3D-HST catalogs. The extraction of fluxes is described in detail in \citet{skelton2014}.   
Fluxes are obtained by measuring the mid-IR intensity using an aperture of 3" on the {\it Spitzer} images.  Contribution from neighboring blended sources in these images are obtained by using a high resolution image.  Fluxes are corrected to account for flux that falls outside the aperture due to the large PSF size of the {\it Spitzer} observations. 
Of the 128 galaxies in 64 merging systems, 102 galaxies have measured {\it Spitzer}/IRAC fluxes with S/N$\ge$3 in the four IRAC bands (44 systems have measurements in both merging galaxies). These mergers have galaxies with projected separations 3.7$-$14.3 kpc  and are in the redshift range 0.5$<$$z$$<$2.4.
In the case of non-mergers, 4493 galaxies  have measured IRAC fluxes and are at 0.31$<$$z$$<$2.5.

\subsection{Radio Data}

We cross-match the merger sample and non-merging galaxies with radio sources  in published radio catalogs. We use the  AEGIS20 Survey (Radio survey of the extended groth strip, \citealt{ivison2007}), the VLA-COSMOS survey \citep{schinnerer2007}, the VLA 1.4 GHz survey of the Extended {\it Chandra} Deep Field-South \citep{miller2008},  VLA 1.4 GHz observations of GOODS-North field \citep{morrison2010}, and 
the ZFOURGE catalogue of AGN candidates  \citep{cowley2016}.

Since the resolution of  the radio observations at 1.4 GHz ranges from 1\farcs7-3\farcs8, we  match merging and non-merging galaxies with the closest radio source at a separation that depends on the beamsize. 
  Of the 64 merging systems, 5 (7.8$\pm^{5.3}_{2.1}$\%) are matched with  a radio source. These systems are at redshift 0.70$\leq$$z$$\leq$2.11 and have galaxies with projected separations 6.9$-$12.2  kpc (0\farcs83-1\farcs5),  therefore we are not able to  identify which of the merging galaxies is the source of radio emission. 
 The matched radio sources have  fluxes in the range $f_{\rm 1.4GHz}$=0.03$-$0.26  mJy which correspond to luminosities $L_{\rm 1.4GHz}$=4.1$\times$10$^{39}-$8.4$\times$10$^{40}$ erg s$^{-1}$,
 and the separation between the merging and the radio sources ranges  0\farcs14$-$1\farcs54.
 In the case of non-merging galaxies, 237 (4.1$\pm$0.3\%) are identified with a radio source.   The radio sources have fluxes in the range $f_{\rm 1.4 GHz}$=0.02$-$16.9  mJy,  redshifts 0.32$\leq$$z$$\leq$2.49, luminosities  $L_{\rm 1.4GHz}$=2.9$\times$10$^{38}-$2.8$\times$10$^{42}$ erg s$^{-1}$, and separations between the galaxy and the radio source between 0\farcs04$-$3\farcs70.

%%%%%% FIGURE : venn_diagrams ->Venn_diagram_use_this.ipynb
\begin{figure*}[!htbp]
\begin{center}
\includegraphics[angle=0,scale=0.73]{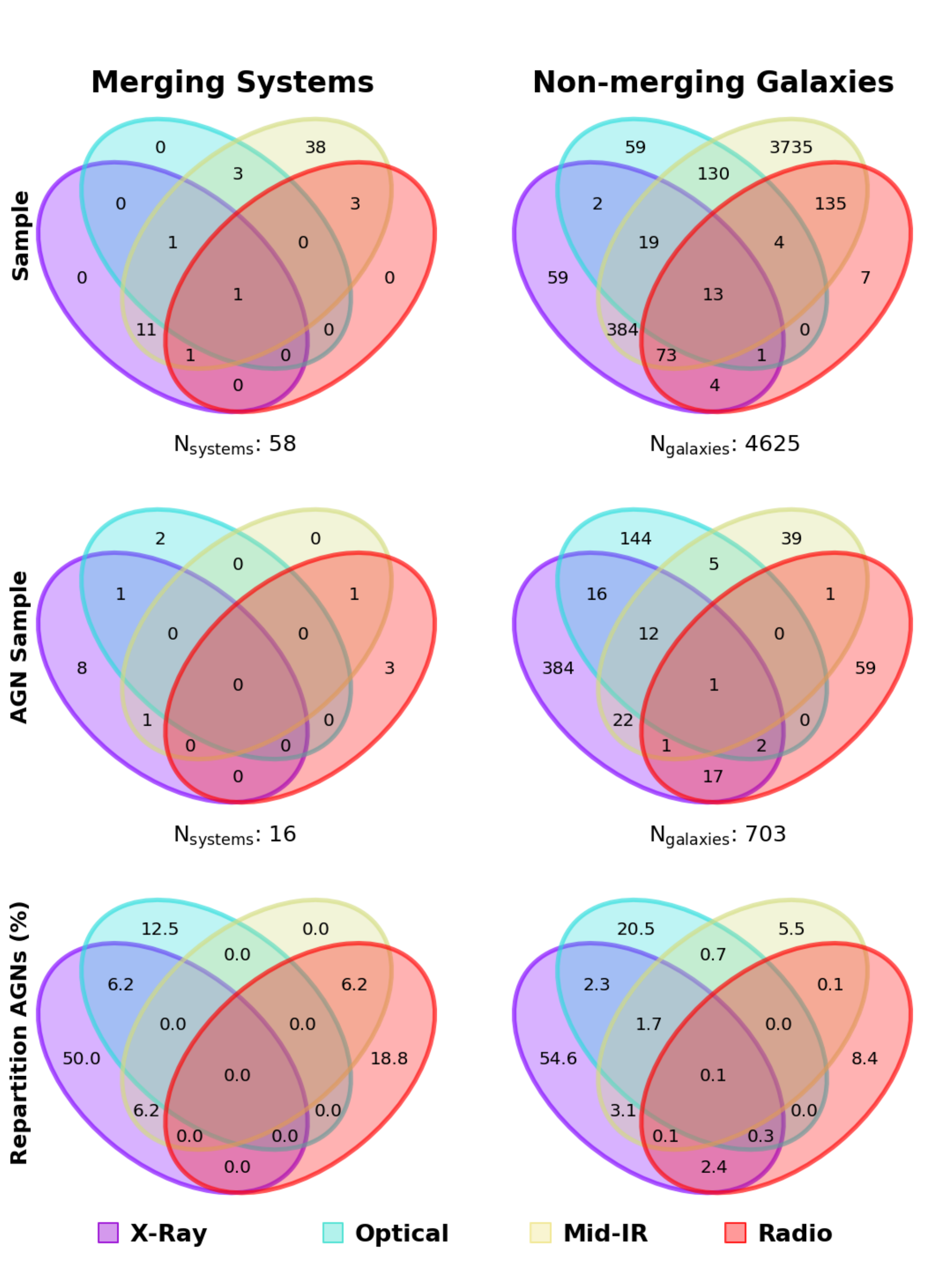}
\caption{ Venn diagrams indicating the combinations in the selection of sources at different wavelengths.  The overlapping regions between samples correspond to the relative number of sources selected in multiple wavelengths. 
Left and right panels present the values for mergers and non-mergers, respectively. 
For simplification, in mergers we include the distribution of merging systems because we are not able to resolve  which of the merging galaxies in  X-ray- and radio-selected AGNs is the AGN source (Sections \S\ref{sel_xray_agn}  and \S\ref{sel_radio_agn}). For non-mergers, we use the values of the selected galaxies.
Top panels correspond to the number of merging systems and non-merging galaxies with measurements to find potential AGNs, i.e. they have either  a matched X-ray or a radio source,  [OIII] and H$\beta$ fluxes, or  IRAC colors in the four bands (Section \S\ref{sec_data_sel}). 
Middle and bottom panels show the number of sources and the repartition in percentage of selected AGNs at the  different wavelengths, respectively (Section \S\ref{sec_agn_selection}).  
\label{fig_venn}}
\end{center}
\end{figure*}

\section{AGN Identification} \label{sec_agn_selection}

AGNs are unique sources 
which manifest  their activity over a wide range of frequencies, due to the various processes that take place in them.  Different methods are employed to identify AGNs at different wavelengths. In  this section, we describe the selection of AGN candidates using techniques in the  X-ray, optical, mid-IR, and 
radio wavelengths and apply them to find potential AGNs in the samples of mergers and non-mergers.

\vspace{1cm}

\subsection{X-ray AGN Selection} \label{sel_xray_agn}

To select X-ray AGNs, we use the technique described in \citet{szokoly2004} and \citet{cowley2016}. This technique applies restrictions on the X-ray luminosity and the hardness ratio HR, which is defined as the normalised difference of counts in the soft and hard X-ray bands $\left(\frac{hard-soft}{hard+soft} \right)$. X-ray AGNs are selected using:
 
\begin{eqnarray}
L_{x} \geq 10^{41}~ {\rm erg~s}^{-1}\hspace{0.2cm}  \& \hspace{0.2cm}  {\rm HR} > -0.2 \\
L_{x} \geq 10^{42}~ {\rm erg~s}^{-1}\hspace{0.2cm}  \& \hspace{0.2cm}  {\rm HR} \leq -0.2 \\
L_{x} \geq 10^{42}~ {\rm erg~s}^{-1}\hspace{0.2cm}  {\rm if} \hspace{0.2cm}  {\rm HR=none.}
\end{eqnarray}

The X-ray luminosity is obtained from 
\begin{eqnarray}
L_{x} [{\rm erg~s}^{-1}] & = & 4 \pi d_{l}^{2} (1+z)^{\Gamma -2} f_{x} ,
\end{eqnarray}
where $f_{x}$ is the measured X-ray flux in erg s$^{-1}$ cm$^{-2}$, $d_{l}$ is the luminosity distance in cm,  and $\Gamma$ is the photon index of the X-ray spectrum. We use $\Gamma$=1.4, a typical value of the photon index found in galaxies \citep{cowley2016}.
 For the redshift $z$, we use the redshift values of the matched merging systems and non-merging galaxies. 
Based on this selection, we find that 10 of the  14 merging systems with a matched X-ray source is a X-ray AGN.  All of these systems have HR measurements (in the range -0.34$<$HR$<$0.39), are at redshifts 0.68$<$$z$$<$2.13, have fluxes    $f_{x}$=7.9$\times$10$^{-17}-$ 2.4$\times$10$^{-14}$ erg s$^{-1}$ cm$^{-2}$, and luminosities  $L_{x}$=1.1$\times$10$^{42}-$ 3.9$\times$10$^{44}$ erg s$^{-1}$. 
In the case of non-mergers, we find that 455 of the 555 X-ray matched sources are  AGNs. These AGNs are in the redshift range 0.34$<$$z$$<$2.49, have fluxes $f_{x}$=1.9$\times$10$^{-17}-$ 1.3$\times$10$^{-13}$ erg s$^{-1}$ cm$^{-2}$, and luminosities  $L_{x}$=1.0$\times$10$^{41}-$ 5.7$\times$10$^{44}$ erg s$^{-1}$.

\subsection{Optical line AGN Selection } \label{sec_opt_agn}

We use the method described in \citet{juneau2014} to find AGNs using both the [OIII] and H$\beta$ optical lines.  For galaxies with masses $\log$(M$_{\star}$/M$_{\odot}$)$\ge$10, AGNs are selected as galaxies in which

\begin{eqnarray}
\log\left(  \frac{f_{\rm [OIII]}}{f_{\rm H\beta}} \right) & > &  410.24 - 109.33 \log  \left[\frac{M_{\star}}{M_{\odot}} \right] \nonumber \\
                                              &+  & 9.7 \log^{2} \left[\frac{M_{\star}}{M_{\odot}} \right]  -0.28  \log^{3} \left[ \frac{M_{\star}}{M_{\odot}} \right] 
\end{eqnarray}
where $f_{\rm [OIII]}/f_{\rm H\beta}$  is the ratio of the [OIII] and H$\beta$ line fluxes and M$_{\star}$ is the stellar mass of the galaxies.  This restriction selects galaxies containing AGNs with a probability  between 60 to 85\%. 
Based on this selection, we find that  4 of the 6 galaxies in mergers with optical lines are optically selected AGNs. These galaxies are at $1.68<z<2.1$ and are in mergers with separations 4.9-13.8 kpc.  Two AGNs are found in one system, i.e. it is a dual AGN system  (two AGNs separated by a closed projected separation \citep[e.g.][]{gross2019, rubinur2019, solanes2019}.
  In the case of non-mergers, 180 of the 228 galaxies with [OIII] and H$_{\beta}$ measurements are AGNs and are in the redshift range 1.4$<$$z$$<$2.3.

\subsection{Mid-Infrared AGN Selection}

We use the  fluxes $f_{3.6 \mu m}$, $f_{4.5 \mu m}$, $f_{5.8 \mu m}$, and $f_{8.0 \mu m}$  in the four IRAC bands to identify mid-IR selected AGNs using the description indicated in \citet{donley2012}.   
mid-IR AGNs are those that follow these five conditions  (valid for $z<2.7$):
\begin{enumerate}
\item   $f_{4.5 \mu m} > f_{3.6 \mu m}$,  $f_{5.8 \mu m}  > f_{4.5 \mu m}$,  and $f_{8.0 \mu m} > f_{5.8 \mu m}$. 
\item  $\log_{10}\left(\frac{f_{5.8 \mu m}}{f_{3.6 \mu m}}\right) \ge 0.08$
\item  $\log_{10}\left(\frac{f_{8.0 \mu m}}{f_{4.5 \mu m}}\right) \ge 0.15$
\item $\log_{10}\left(\frac{f_{8.0 \mu m}}{f_{4.5 \mu m}}\right) \ge \left[1.21 \times  \log_{10} \left(\frac{f_{5.8 \mu m}}{f_{3.6 \mu m}}\right) \right]-0.27$
\item  $\log_{10}\left(\frac{f_{8.0 \mu m}}{f_{4.5 \mu m}}\right) \le \left[1.21 \times  \log_{10} \left(\frac{f_{5.8 \mu m}}{f_{3.6 \mu m}}\right) \right]+0.27$
\end{enumerate}

Of the 102 galaxies  with  measured IRAC fluxes, we find that 2 galaxies are mid-IR selected AGNs.   The galaxies have projected separations  of 7.4 and 12.2 kpc and  redshifts $z$=1.9 and 2.11.
For non-mergers, 81 of the 4493 galaxies with measured mid-IR fluxes are AGNs and are in the redshift range 0.44$<$$z$$<$2.49.

\subsection{Radio AGN Selection} \label{sel_radio_agn}

Radio AGNs are identified as those sources that have an excess of the star formation rate in the radio compared to the star formation rates obtained in the Infrared and UV wavelengths.  
When this Radio-AGN activity index  is above 3 (SFR$_{\rm Radio}$/SFR$_{\rm UV+IR}>$3), galaxies are classified as radio AGNs \citep{rees2016}. 
  The UV and IR star formation rates correspond to the unobscured and obscured star formation activity in the galaxies, respectively. 
  The use of the SFR$_{\rm UV+IR}$ and not only the SFR$_{\rm IR}$ in the Radio-AGN activity index avoids the misclassification as a radio-AGN of radio, low dust, star-forming galaxies, which can produce an excess in the SFR$_{\rm Radio}$ if they are compared only with the obscured star formation. 
  
  The 10 galaxies in the 5 merging systems with a matched radio source have
   SFRs obtained from the combination of rest-frame UV emission and mid-IR pho\-to\-me\-try obtained from {\it Spitzer}/MIPS imaging. 
  For the 237 non-merging galaxies with a matched radio source, 217 have SFRs from UV+IR and 20 from modeling their SEDs
  
  The SFR$_{\rm Radio}$ is measured from the rest-frame radio luminosity $L_{\rm Radio}$ using the description indicated in \citet{cowley2016}:
 \begin{equation}
 \rm SFR_{\rm Radio} [\rm M_{\odot} yr^{-1}]=3.18 \times 10^{-22} (L_{\rm Radio}/ WHz^{-1}).
 \end{equation} 
 
 The rest-frame radio luminosity is given by 
 \begin{equation}
 L_{\rm Radio} [{\rm W Hz}^{-1}] = 4 \pi d_{l}^{2} (1+z)^{-(\alpha +1)} f_{\rm Radio}
 \end{equation}
 where $d_{l}$ is the luminosity distance in cm, $f_{\rm Radio}$ is the measured radio flux in W m$^{-2}$ Hz$^{-1}$, and $\alpha$ is the radio spectral index. In this work, we use the standard $\alpha=-0.7$ value\footnote{ Typical of optically thin synchrotron emission.}.  The redshift $z$ is obtained from redshift values of the matched mergers and non-merging galaxies.
  
  We measure the radio-AGN activity index in non-mergers by measuring the ratio of the SFR$_{\rm Radio}$ of the matched radio source over the measured SFR value (SFR$_{\rm UV+IR}$ or SFR$_{\rm SED}$). In the case of mergers, since we do not know which of the merging galaxies is the radio-AGN source,  we assume that the galaxy with the highest SFR$_{\rm UV+IR}$ is the radio source. 
   Therefore, the radio-AGN activity index in mergers is a lower limit.  
  
 We find that 4 of the 5 radio-matched merging systems are radio-AGNs. These AGNs have radio luminosities  L$_{\rm 1.4 GHz}$=4.1$\times 10^{39}-$8.4$\times 10^{40}$ erg s$^{-1}$, are in systems in which the merging galaxies have projected separations 6.9$-$10.7 kpc (0\farcs83-1\farcs5), and are at redshifts 0.70$<$$z$$<$2.11.
 For non-mergers, 81 of the 237 matched galaxies are radio-AGNs. The AGNs have luminosities $L_{\rm 1.4 GHz}$=3.3$\times 10^{38}-$ 2.8$\times 10^{42}$ erg s$^{-1}$ and are in the redshift range 0.32$<$$z$$<$2.48.

 \subsection{Combined Selection}\label{sec_combined}
 
 We combined the results of the previous AGN selections to analyze how many merging systems and non-merging galaxies have
 single or multiple AGN types.
 Of the 64 merging systems, 58 systems (105 galaxies) have either a matched X-ray source, IRAC colors in the four bands,  [OIII] and H$\beta$ optical lines and/or  a matched radio source. We find that 10 and 4 systems are identified with an X-ray and radio AGN, respectively and 4 and 2  individual merging galaxies (in 3 and 2 merging systems) are optical- and mid-IR-selected AGNs, respectively.  There are 13 systems identified with only one AGN type while 3 systems are identified with multiple AGN types.
  In the case of non-mergers, 4625 galaxies have measurements to find potential AGNs.  We find that 703 galaxies are identified as AGNs with  
  455, 180, 81 and 81 identified as X-ray, optical, mid-IR, and radio AGN, respectively. 
  Of these 703 AGNs,  626 galaxies are identified with one AGN type and 77 with multiple AGN types. 
 Figure \ref{fig_venn} presents Venn diagrams indicating the combination of the results of the matched sources and AGN galaxies selected in the different wavelength ranges\footnote{Venn diagrams were obtained by using {\sc pyvenn} (https://github.com/LankyCyril/pyvenn/blob/master/pyvenn-demo.ipynb)}.

 %%%%%    RESULTS   %%%%%  
\section{Results} \label{sec_results}

 \subsection{Total AGN Fraction}
  
 \subsubsection{Non-mergers}
 The total AGN fraction for non-mergers can be  obtained from   $f_{\rm AGN}=N_{\rm AGN}/ N_{\rm T}$, which is the ratio of the number of galaxies that are AGNs ($N_{\rm AGN}$=703) and the number of galaxies with measurements to find potential AGNs ($N_{\rm T}$=4625, Section \S \ref{sec_combined}).  In this case, the  total AGN fraction in non-mergers is $f_{\rm AGN}$=15.2$\pm$0.6\%.  Uncertainties are obtained from 10000 Monte Carlo realizations, where for each matched source, we draw a flux from the measured fluxes ($f_{\rm x}$, $f_{\rm [OIII]}$, $f_{\rm H\beta}$,  $f_{3.6 \mu m}$,$f_{4.5 \mu m}$, $f_{5.8 \mu m}$, $f_{8.0 \mu m}$, $f_{\rm 1.4GHz}$) and perturb them with the errors of the flux measurements.  For each realization, we calculate how many galaxies meet the criteria to be classified as AGNs and calculate the AGN fraction. The uncertainty is calculated from the standard deviation  of the resulting fractions and then added in quadratures with 1$\sigma$ errors  obtained from Poisson statistics \citep{gehrels1986}.
 Due to the  varying sensitivity of the {\it Chandra} observations, the denominator  in the fraction  needs to be corrected to account for the effective number of galaxies for which an X-ray-selected AGN could have been  detected (as in \citealt{silverman2011}).

  The corrected AGN fraction is obtained from:
 \begin{eqnarray}
 f_{\rm AGN-c} & = &  \hspace{0.5cm}  f_{\rm nonXR}   \hspace{0.5cm} +  \hspace{1.0cm}   f_{\rm XR}  \nonumber \\
  &= &    \frac{N_{\rm AGN-nonXR}  }{N_{\rm T} }  +    \sum_{i=1}^{N_{\rm AGN-XR}} \frac{1}{N_{\rm eff-i} + N_{\rm nonXR}}  \hspace{0.5cm} 
 \end{eqnarray}
 
\begin{figure}[!htbp]
\begin{center}
\includegraphics[angle=0,scale=0.4]{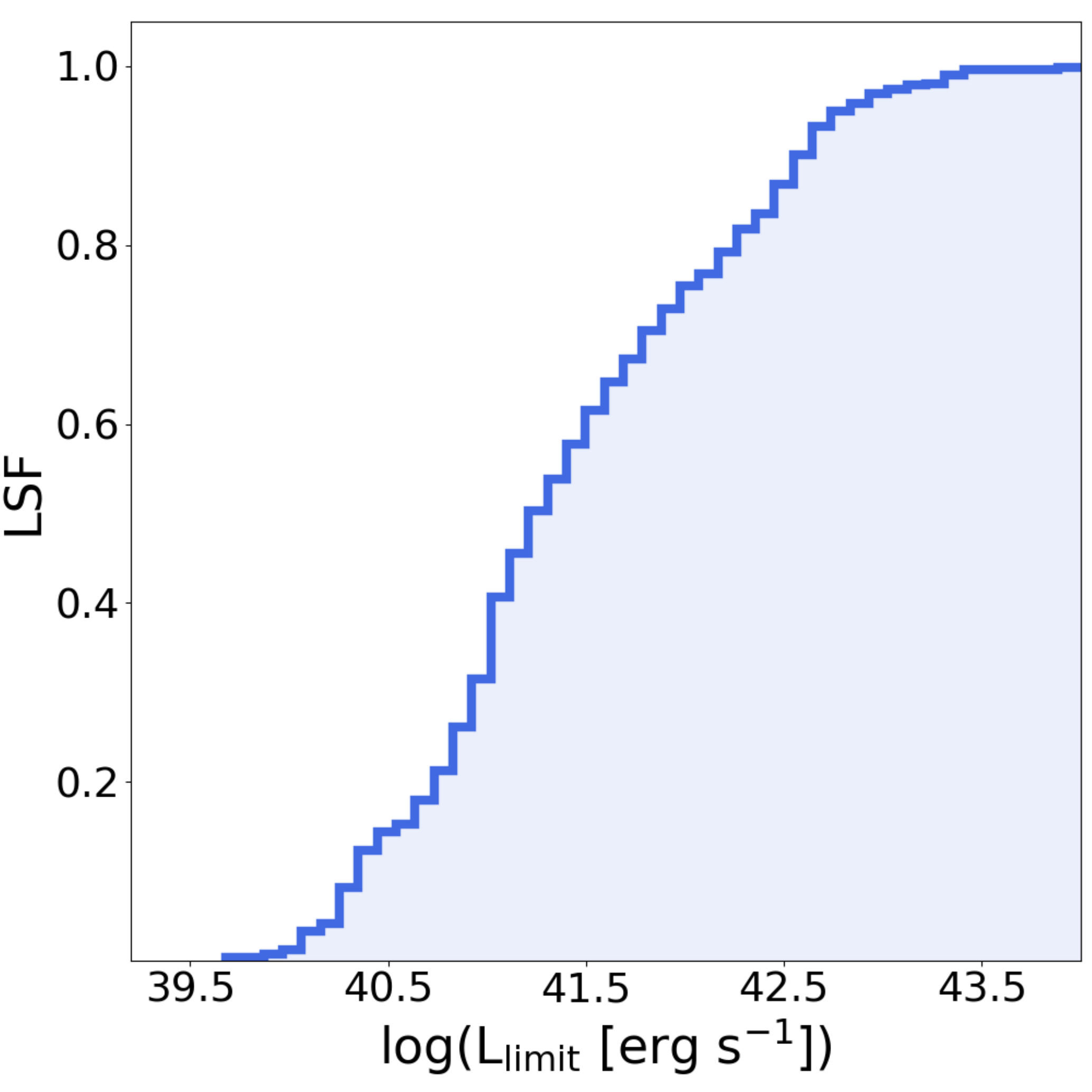}
\caption{Luminosity sensitivity function (LSF) as a function of X-ray luminosity.  This function indicates the fraction of galaxies where an AGN could have been detected above a given X-ray luminosity. 
\label{fig_LSF}}
\end{center}
\end{figure}

 The fraction  $f_{\rm nonXR}$\footnote{We do not apply any correction.} corresponds to  the ratio of the number of AGNs not selected in X-ray at all ($N_{\rm AGN-nonXR}$=248)  and the total number of galaxies with measurements to find potential AGNs ($N_{\rm T}$=4625).   
 The fraction $f_{\rm XR}$ is the  
 AGN fraction for  X-ray-selected AGNs (even if they  were identified as AGNs in other wavelengths). 
 In this case, for each X-ray-selected AGN (with a total of $N_{\rm AGN-XR}$=455 galaxies), we  measure the number of effective galaxies in which we could have detected an X-ray AGN  ($N_{\rm eff-i}$) and then add the number of galaxies with measurements to find potential AGNs in optical, mid-IR, and radio only ($N_{\rm nonXR}$=4070).
To calculate $N_{\rm eff-i}$,    we produce the luminosity sensitivity function  (LSF, \citealt{birchall2020}), which is the distribution of the fraction of galaxies where an AGN could have been detected above a given X-ray luminosity.  To obtain this function, we  use all the sources in our sample (mergers and non-mergers)  that are spatially matched with an X-ray source (575 galaxies).  For each of these matched galaxies, we assign a flux limit which correspond to the  flux limit of  the survey where the galaxy was  extracted. 
Using the redshifts of the matched galaxies in our sample, the flux limit is converted into a luminosity limit. The cumulative histogram  as a function of the luminosity limits of these sources is normalized by the size of the sample of matched galaxies (Fig. \ref{fig_LSF}).

%%%%%%. FIGURE   cumm_distrib_lum.py 

 %%%%%%%%.  FIGURE   plot_frac_general.py 
 \begin{figure*}[!htbp]
\begin{center}
\includegraphics[angle=0,scale=0.6]{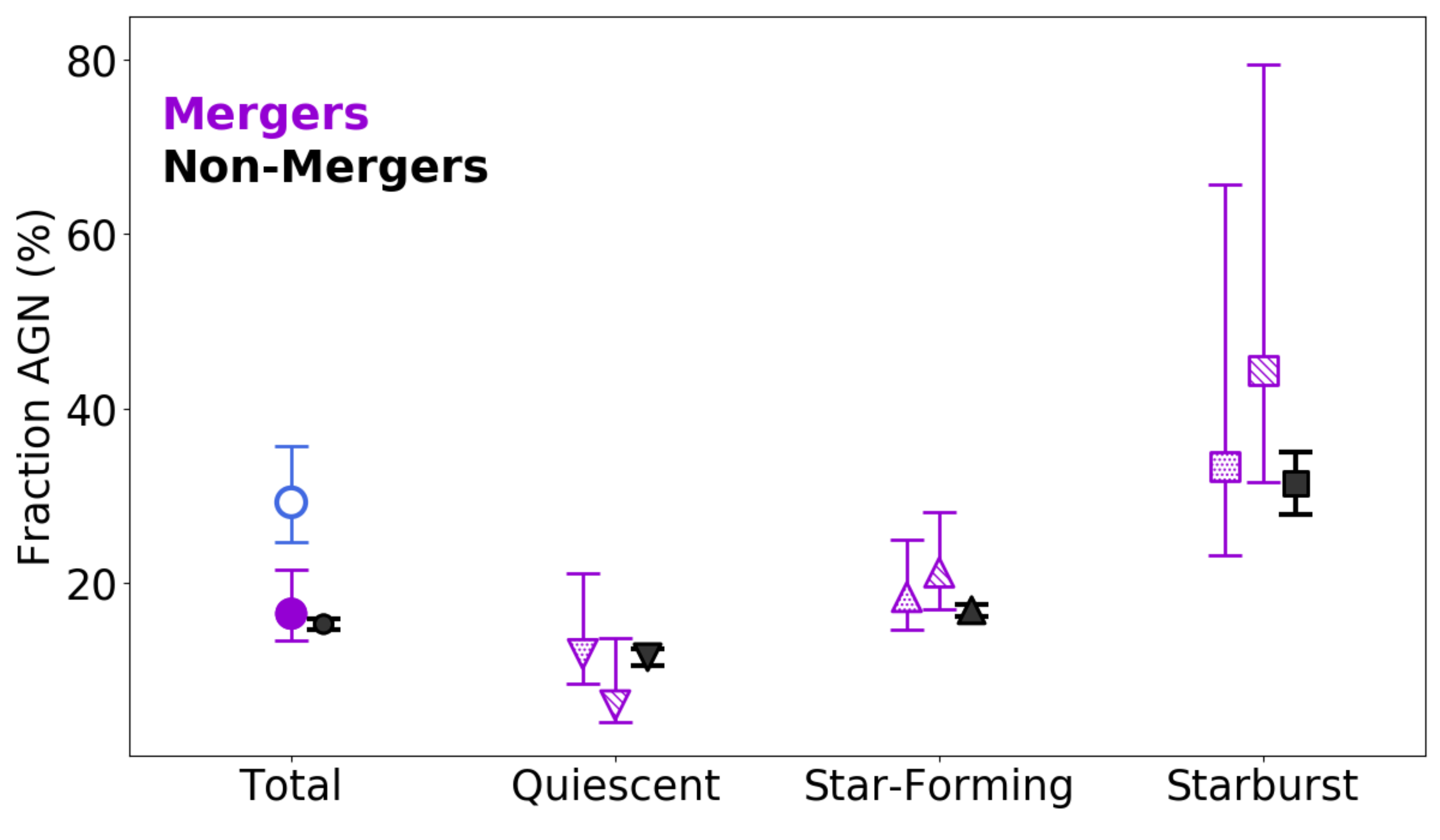}
\caption{Fraction  of AGNs in mergers (purple symbols) and in non-merging galaxies (black symbols). 
Circles show the total AGN fraction.  The filled purple and blue open circles indicate the total fraction of AGNs in mergers assuming that, when the AGN galaxy is not resolved,  one and two merging galaxies are AGNs, respectively. 
The  downward and upward triangles correspond to the AGN fraction for  star-forming and quiescent galaxies, respectively. The dashed and dotted filling  correspond to the assumption that, in mixed mergers, the emitter of an unresolved AGN is the star-forming and the quiescent galaxy, respectively.  
The squares corresponds to the fraction of AGNs in galaxies that are starbursts (i.e., galaxies that lie 0.5 dex above the main sequence fit presented in  \citealt{whitaker2014}).  Errors are obtained from Poisson statistic \citep{gehrels1986}.}
\label{fig_frac_agn}
\end{center}
\end{figure*}

For each X-ray-selected AGN with a luminosity $L_{x-i}$,  we use the LSF to measure the effective number of galaxies where the {\it Chandra} observations were sufficiently sensitive to detect an AGN of this luminosity. This effective number is given by   $N_{\rm eff-i}={\rm LSF_{i}} \times N_{\rm XR}$, where $N_{\rm XR}$ is the number of non-merging galaxies with measurements to find potential X-ray-selected AGNs  ($N_{\rm XR}$=555) and LSF$_{\rm i}$  is the value of the luminosity sensitivity function  at a luminosity $L_{\rm lim}$=$L_{x-i}$. 
For instance, for an X-ray-selected  AGN with $\log(L_{x} {\rm [erg~ s}^{-1}])$$\geq$43.5, the effective number of galaxies is  $N_{\rm eff-i}$=1.0$\times$555=555
, while for an  AGN with luminosity $\log(L_{\rm x} [{\rm erg~s}^{-1}])$=41.5 the effective number is   $N_{\rm eff-i}$=0.61$\times$555=338.5. 
After measuring  $N_{\rm eff-i}$ for all the X-ray-selected AGNs, we sum them up and find a total corrected AGN fraction of  $f_{\rm AGN-c}$= 15.4$\pm$0.6\% (black  filled circle in Fig. \ref{fig_frac_agn}). This fraction is not significantly different from the non-corrected AGN fraction $f_{\rm AGN}$=15.2$\pm$0.6\%.

 \subsubsection{Mergers}
 
 For the AGN fraction in  mergers, in addition to correcting for the sensitivity variation of the X-ray observations, we need to take into account that  for  X-ray and radio sources, we are not able to  resolve which of the merging galaxies in a system is the emitter.  
 All the merging galaxies used in this work have either  resolved mid-IR colors or  [OIII]+H$\beta$-detected lines, thus the number of merging sources with measurements to find potential AGNs is $N_{\rm T}$=105 (in 58 merging systems).  The number of AGNs in mergers is measured in the following way.  First, when mergers are identified solely with an X-ray or a radio AGN, we assume that one of the merging galaxies is the AGN source. If the X-ray/radio AGN is also an optical or mid-IR selected AGN, we assume that the resolved optical/mid-IR-selected AGN is also the X-ray/Radio AGN.  For mergers identified with an optical or mid-IR-selected AGN, the number of AGNs is just the number of individual galaxies identified with those AGNs.  
 With this assumption, we find that 17 galaxies (in 16 merging systems) are active nuclei. 
 We find $N_{\rm AGN-nonXR}$=6, $N_{\rm nonXR}$=77, $N_{\rm AGN-XR}$=11,  and $N_{\rm XR}$=28 individual merging galaxies. We use the same LSF used for non-merging galaxies to measure the number of effective galaxies $N_{\rm eff-i}$. The corrected AGN fraction for mergers is $f_{\rm AGN-c}$=16.4$\pm ^{5.0}_{3.1}$\% (purple filled circle in Fig. \ref{fig_frac_agn}). Without correction for the X-ray sensitivities, the AGN fraction would be  $f_{\rm AGN}$=16.2$\pm ^{4.9}_{3.0}$\% which is not significantly different of the corrected fraction. 
 
 The second way to measure the AGN fraction in mergers is to assume that, when a system is identified with an X-ray or a radio AGN, the system is a dual AGN,  even though one of the merging galaxies is also  identified with an optical or mid-IR-selected AGN. Of the 16 merging systems with AGNs,  1 system contains two optically identifeid AGNs (also classified as an X-ray AGN), 2 systems contain only one AGN in each of them (in both cases the AGNs are identified only in optical),  and 13  systems are identified with an X-ray or radio AGN (some of them are AGNs  identified also in other wavelengths).
 With this assumption, we find that 30 galaxies are active nuclei and the AGN fraction in mergers is  $f_{\rm AGN-c}$=29.2$\pm ^{6.5} _{4.5}$\% (open  blue circle in Fig. \ref{fig_frac_agn}). Therefore, the total fraction of AGNs in mergers ranges between this number and the fraction found assuming that one of the merging galaxies is the source of unresolved emission of the X-ray/radio AGNs. However, the general fraction of dual AGNs with respect to the total AGN population has been found to be low, with a fraction at most of 20--30\%  \citep{rosario2011, koss2012, muller2015, volonteri2016, capelo2017, solanes2019, rosasguevara2019, imanishi2020, silverman2020}.  Therefore, the total AGN fraction should be closer to the lower value. 
  We find that one of the 16 merging systems containing AGNs is a dual AGN (see \S \ref{sec_opt_agn}) and 
  it is possible that there are more of them in this mergers sample. Higher resolution observations are needed to identify if the other 11 merging systems containing X-ray and/or radio AGNs are dual AGNs.

\subsubsection{Comparison AGN Fraction  between Mergers and Non-mergers}
 
\begin{figure}[!htbp]
\begin{center}
\includegraphics[angle=0,scale=0.40]{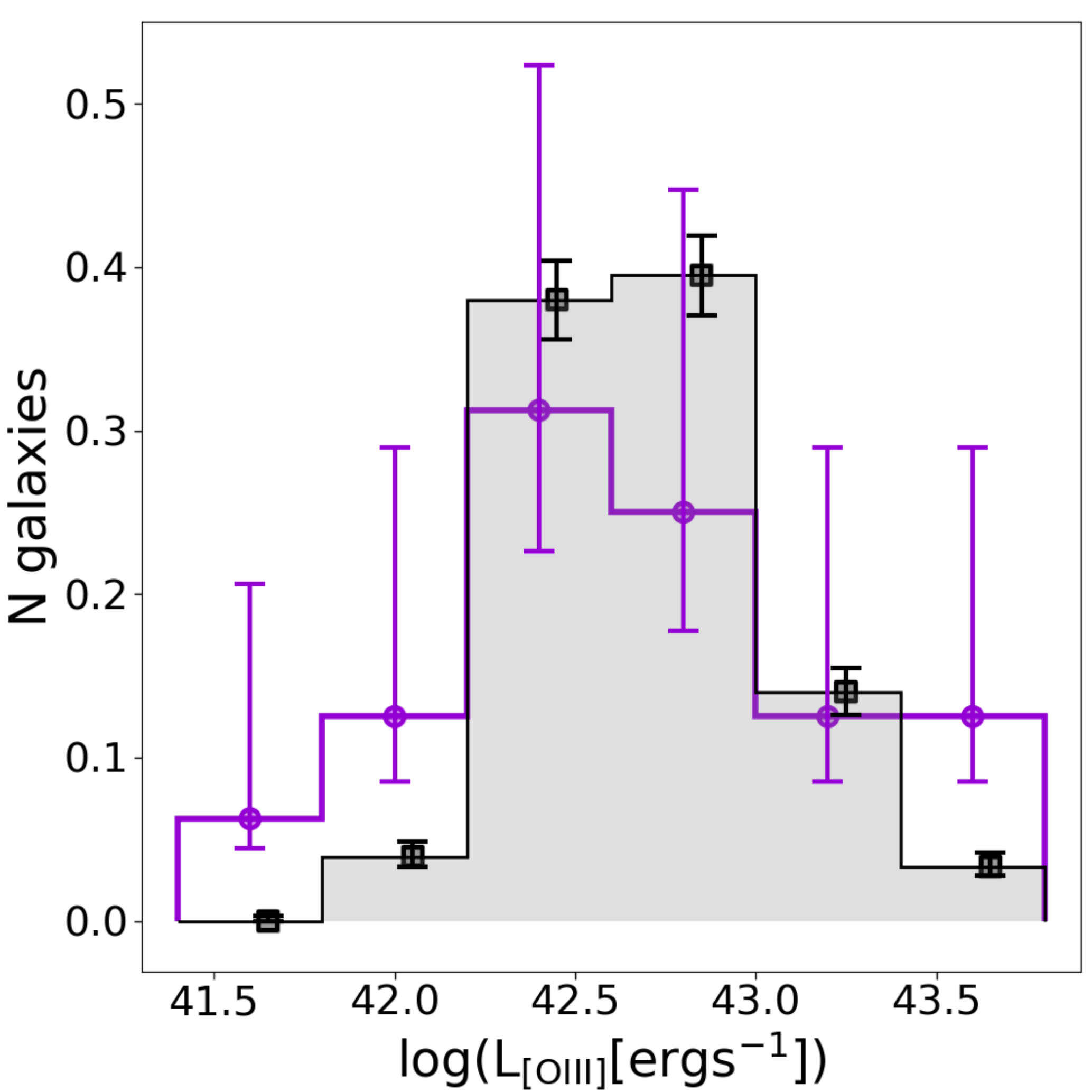}
\caption{ Normalized distribution of the [OIII] luminosity (addition of the values is 1) for merging (purple histogram) and non-merging galaxies (filled grey histogram).  Error bars  are obtained from Poisson statistic.  
\label{fig_dist_oiii}}
\end{center}
\end{figure}

  We find a similar AGN fraction in merging (16.4$\pm^{5.0}_{3.1}$\%) and non-merging galaxies (15.4$\pm$0.6\%) when assuming that the source of the unresolved AGNs (X-ray or radio) is only one  of the merging galaxies in a system.  Since the [OIII] luminosity itself is an indicator of AGN activity at high luminosities, we compare the normalized  distribution of the [OIII] luminosity  between mergers and non-mergers (Fig. \ref{fig_dist_oiii}).  An excess in this distribution would suggest higher AGN activity.
  We select all merging and non-merging galaxies with [OIII] measurements and S/N$\ge$3, even if they do not have H$\beta$ measurements as used in the data selection  presented in Section  \S \ref{optical_data}.  We find 13 and 730 merging and non-merging galaxies with [OIII] measurements, respectively. 
  We perform a Kolmogorov-Smirnov (K-S) test between the distributions of the merging and non-merging samples to 
   quantify whether the two samples are consistent with coming from the same parent population. We find a P-value (probability of the plausibility of the null hypothesis) of $P$=0.32, therefore the two populations are indistinguishable at more than 3$\sigma$ (since $P>0.003$\footnote{We also perform this analysis in different bins of redshifts, star formation rates, and stellar masses and find no significant difference.}), in agreement with the lack of AGN excess  in mergers found using the multiwavelength AGN selection.

  \vspace{0.5cm}
 {\it  In the following results, we use  the assumption that only one of the merging galaxies is the  X-ray or the radio-selected AGN.}

%%%%%%%%%%%%%.   FIGURE   plot_distrib_oiii.py

\subsection{AGN Fraction for Different Merger/Galaxy Types}

As indicated in Section  \S \ref{sec_merg_sample}, we separate galaxies into star-forming (dusty or unobscured) and quiescent  using the rest-frame UVJ colors of the galaxies. Using this classification, we divide mergers into wet (two star-forming), mixed (one star-forming and one quiescent), and dry (two quiescent).  
The downward and upward triangles in Figure \ref{fig_frac_agn} present the AGN fraction for quiescent ($f_{\rm Q}$) and star-forming galaxies ($f_{\rm SF}$), respectively. 
For non-mergers,   the fractions are obtained from the ratio of number of AGNs ($N_{\rm A}$) over the total number of galaxies ($N_{\rm t}$) with measurements to find potential AGNs of their respectively galaxy type ($f_{\rm Q}=N_{\rm A_{Q}}/N_{\rm t_{Q}}$ and $f_{\rm SF} =N_{\rm A_{SF}}/N_{\rm t_{SF}}$).  The AGN fractions in non-mergers are  $f_{\rm Q}$=11.4$\pm$0.9\% and  $f_{\rm SF}$=16.8$\pm$0.7\%.\footnote{Since the total AGN fractions corrected and noncorrected by the variation of X-ray sensitivity in mergers and non-mergers are significantly the same, we do not apply a correction in the measurement of the following fractions.}
For mergers, the AGN fraction for the different galaxy types is not straightforward.  The complication arises  in the case of  a mixed merger identified with a non-resolved  X-ray or radio AGN and with no other resolved AGN selected in optical or mid-IR. In wet and dry mergers, since both merging galaxies are of the same type, the emitter in an unresolved  AGN will be either a star-forming or a quiescent galaxy, respectively.   The AGN fraction for star-forming and quiescent  merging galaxies will be measured in two ways: 
1) the AGNs in mixed mergers arise from  the star-forming galaxies  (dashed purple triangles in Fig. \ref{fig_frac_agn}), which gives an AGN fractions of    $f_{\rm Q}$=5.9$\pm ^{7.8} _{1.9}$\%  and $f_{\rm SF}$= 21.1$\pm ^{7.0} _{4.2}$\% ;
2) the AGNs arise from the quiescent   galaxies (dotted purple triangles in Fig. \ref{fig_frac_agn}),  resulting in AGN fractions of   $f_{\rm Q}$=11.8$\pm ^{9.3} _{3.4}$\% and $f_{\rm SF}$=18.3$\pm ^{6.6} _{3.8}$\%.
 In summary, we find that in non-mergers the AGN fraction in star-forming galaxies is higher than in quiescent galaxies. The same trend is found in mergers, which is more significant when assuming that the unresolved AGN arises from the star-forming galaxy in a mix merger.

 The square symbols in Figure \ref{fig_frac_agn} present the AGN fractions  found for starbursts, i.e., galaxies that lie 0.5 dex above the main sequence fit obtained by \citet{whitaker2014}. For non-mergers, the fraction is obtained from the ratio of the number of starbursts that are AGNs over the total number of starburst galaxies with measurements to find AGNs ($f_{\rm SB}$=$N_{\rm A_{SB}}/N_{\rm t_{SB}}$). We find an AGN fraction in  starbursting non-merging  galaxies of $f_{\rm SB}$=31.4$\pm$3.6\%.  
 For mergers,  we find an  AGN fraction in starbursting merging galaxies of $f_{\rm SB}$=44.4$\pm ^{35.1} _{12.9}$\%(dashed purple square).  However, this value is an upper limit, since we assume that for non-resolved AGNs the starburst is also  the AGN galaxy\footnote{Since a starburst is also a star-forming galaxy, this fraction also corresponds to the fraction obtained when assuming that in an unresolved AGN, the star-forming galaxy is the AGN source.}.   If not, this value drops to $f_{\rm SB}$=33.3$\pm ^{32.4} _{10.2}$\%\footnote{ Same fraction is obtained if we assume the AGN comes from the quiescent galaxy.} (dotted purple square).  In any case, the AGN fraction in starbursts in mergers and non-mergers is higher than the total AGN fraction of the general population.

\begin{figure}[!htbp]
\begin{center}
\includegraphics[angle=0,scale=0.32]{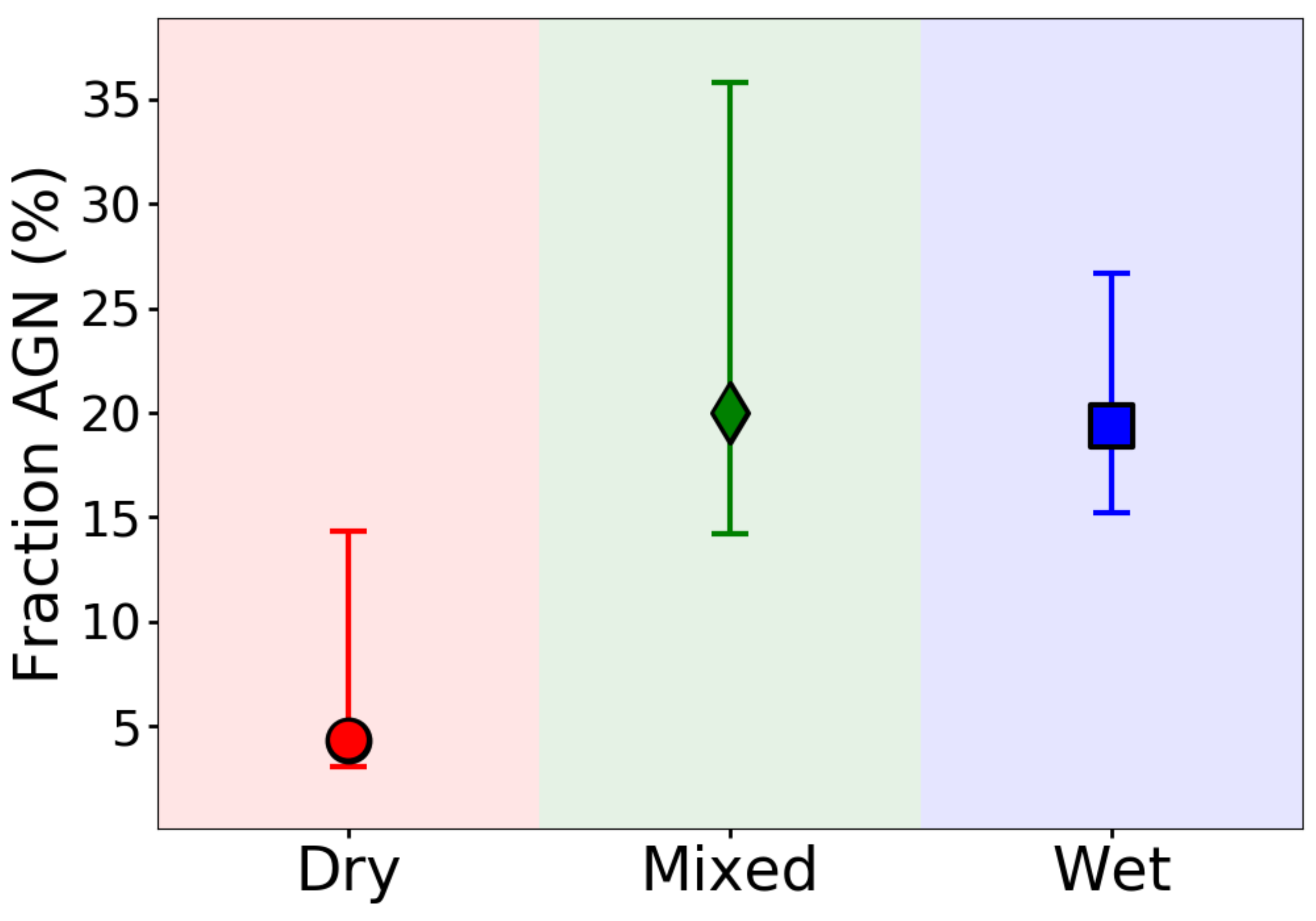}
\caption{Fraction of AGNs in merging systems depending on their type. Blue square, green diamond, and red circle correspond to the AGN fraction in wet (two star-forming galaxies), mixed (one star-forming and one quiescent), and dry mergers (two quiescent galaxies), respectively.  Errors are obtained from Poisson statistic.
\label{fig_frac_mergtype}}
\end{center}
\end{figure}

Figure \ref{fig_frac_mergtype} shows the fraction of AGNs in mergers depending of the merger type. We find AGN fractions of  4.3$\pm ^{10.0} _{1.3}$\%,  20.0$\pm ^{15.8} _{5.8}$\%, and 
19.3$\pm ^{7.3}_{4.1}$\%   for dry, mixed, and wet mergers, respectively.  This results might suggest that the presence of a star-forming galaxy in a merging system increases the incidence of AGNs, although the excess in the fraction in wet and mixed mergers compared to dry mergers is not significant given the associated error bars.

%%%%%%%%%%%%.  FIGURE.  frac_mergtype.py 

  \subsection{Evolution AGN Fraction}
  
\begin{figure*}[!htbp]
\begin{center}
\includegraphics[angle=0,scale=0.61]{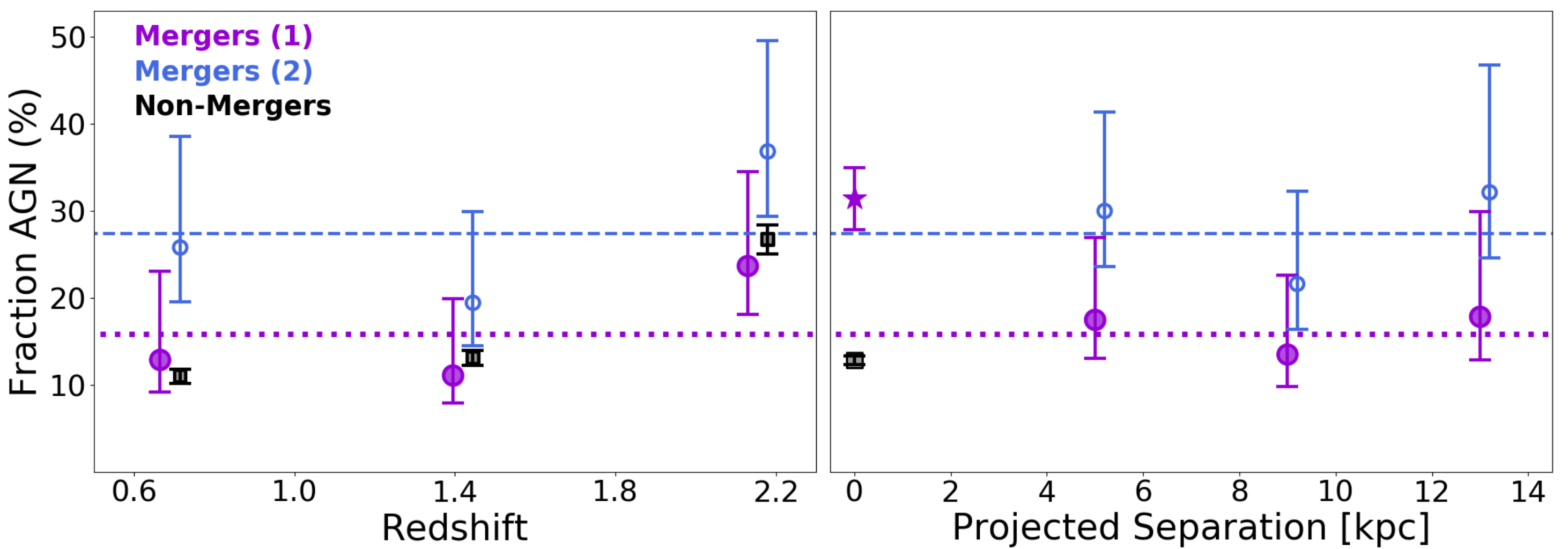}
\caption{ Fraction of AGNs in mergers and non-mergers as a function of redshift (left) and projected separation between the galaxy nuclei (right).  Purple circles correspond to the AGN fraction in mergers assuming that one of the merging galaxies is the unresolved radio or X-ray AGN, while the  small  blue circles correspond to the AGN fraction assuming that both merging galaxies are the AGN source.  Black squares in the left correspond to the AGN fraction as a function of redshift for non-merging galaxies, while the black square in the right correspond to  the total AGN fraction. The purple star correspond to the fraction of AGNs assuming that all the starburst galaxies in our sample of non-mergers are mergers at the coalescence phase.
The dashed and dotted lines indicate the mean of the fraction values obtained for mergers. 
 Errors are obtained from Poisson statistics. 
\label{fig_frac_z_sep}}
\end{center}
\end{figure*}

Figure \ref{fig_frac_z_sep}, left shows the fraction of AGNs in mergers (purple circles) and in non-mergers (black squares) as a function of redshift. The dotted line is  the median of the three fraction values obtained for mergers.  
We find that the total AGN fraction in non-mergers increases with redshift, while for mergers 
it is almost constant at $z$$\lesssim$1.5 and increases at higher redshifts. We also include the evolution with redshift of the AGN fraction assuming that the unresolved AGNs are dual (blue circles in Fig. \ref{fig_frac_z_sep}).

%%%%%%. FIGURES plot_frac_z_sep.py 

Figure \ref{fig_frac_z_sep}, right shows the fraction of AGNs in mergers as a function of the projected separation of the merging galaxies.  The black square shows the total AGN fraction for non-merging galaxies. 
The AGN fraction  in all bins of projected separation are marginally larger than the fraction for  non-mergers but not significantly due to the large error bars.
Note that we are missing the coalescence phase, when the highest black hole accretion activity is expected \citep[e.g.][]{springel2005, hopkins2008, debuhr2011}.  % .  {\bf based on simulaitons reference}
Some of the starburst galaxies in the non-merger sample are possibly mergers at the coalescence phase.  For instance, \citealt{cibinel2019} investigate the fraction of close pairs and morphologically identified mergers on and above the star-forming main sequence at 0.2$<z<$2.0. They found that starburst galaxies are mostly dominated (more than 70\%) by mergers in their late stages.  If for simplicity, we assume that all the starburst in the non-merging sample are mergers at coalescence\footnote{Maintaining the starbursts found in the mergers sample as galaxies in the pre-coalescence phase.} (78 of the 248 matched starbursting non-merging galaxies are AGN), then the fraction of AGNs at the coalescence phase (i.e. projected separation is zero) is 31.4$\pm3.6$\% (purple star in Fig. \ref{fig_frac_z_sep})  and we can see a clear increment in the AGN fraction at projected separations below 10 kpc. With this assumption, 
the total fraction of AGNs in mergers increases from 16.4$\pm^{5.0}_{3.1}$\% to 26.9$\pm$2.8\% while in non-mergers the fraction changes from 15.4$\pm$0.6\% to  14.3$\pm$0.6\%.  Therefore, the  fraction of AGNs in non-mergers remains almost the same as originally measured while in mergers it increases significantly. 
However, contrary to the merger origin of starbursts, there are also findings indicating that high redshift starbursts are possibly the result of galaxies with high levels of cold gas content \citep{scoville2016, oteo2017} rather than mergers at coalescence. 
It is out the scope of this paper to identify which of these starbursts are actually the result of a merger event.

%%%%% FIGURE. plot_main_sequence_bhar.py 
\begin{figure*}[!htbp]
\begin{center}
\includegraphics[angle=0,scale=0.69]{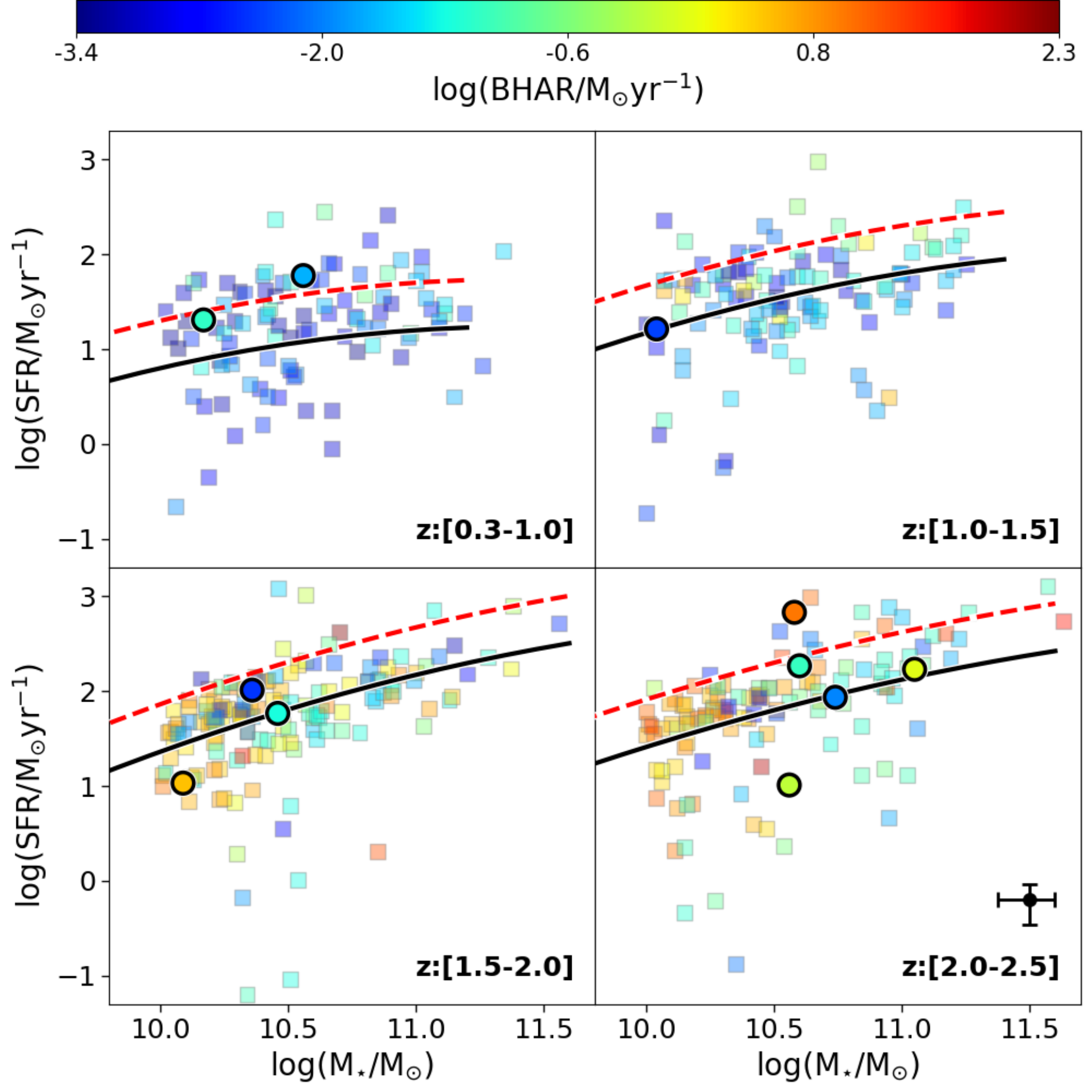}
\caption{Star formation rate as a function of stellar mass  for star-forming merging and non-merging galaxies  in different redshift bins. Galaxies are color-coded by the intensity of black hole accretion rate. Circles are the position of merging galaxies (11 of the 12 merging galaxies with X-ray and/or [OIII] emission are star-forming) while small squares are the position of non-merging star-forming galaxies. The black curves show the main sequence fit obtained by \citet{whitaker2014} while the red dashed curve indicate the minimum position for which galaxies are considered as starbursts.  Median of the uncertainties in star-formation rate and stellar mass are shown in the bottom right panel.
\label{fig_main_seq_bhar}}
\end{center}
\end{figure*}

 \vspace{1cm}

\subsection{Relation between BHAR and SFR}

To analyze if the black hole accretion rate (BHAR) is related to the star-formation rate and/or the stellar mass in the galaxies, we use only X-ray and optically-selected AGNs. With the X-ray and [OIII] luminosities of these AGNs, we can estimate their bolometric luminosity $L_{\rm bol}$ and  the BHAR from $L_{\rm bol}$=$\epsilon \dot m c^{-2}$, where $\epsilon$ is the  matter-to-radiation efficiency conversion, $\dot  m$ is the black hole accretion rate, and  $c$ is the light speed. 
Of the 703 non-merging galaxies with AGNs, 596 are  X-ray and/or optically-selected AGNs  (we use  447  and 149 galaxies with  X-ray and optical emission, respectively).  For mergers,  12 systems are X-ray-selected AGNs while 2 merging galaxies are AGNs selected from their [OIII]+H$\beta$ emission. When a merging system or a merging galaxy is selected as  an optical and X-ray-selected AGN, we will use the X-ray emission to measure the bolometric luminosity, because the [OIII] emission is uncorrected by dust attenuation. To perform this correction, it is necessary a measurement of the Balmer decrement H$\alpha$/H$\beta$. However, these lines are not available simultaneously for our sources. 
Bolometric luminosities can be estimated assuming
 \begin{eqnarray}
 L_{\rm bol} & = &22.4~ L_{\rm X}, \\
 L_{\rm bol} & = & 3500~ L_{\rm [OIII]}, 
 \end{eqnarray}
where $L_{\rm X}$ and $L_{\rm [OIII]}$ are the X-ray and [OIII] luminosities, respectively. In X-ray, the bolometric luminosity is obtained by assuming a conversion factor  $k_{bol}$=22.4, which is the median value obtained for local AGNs with $L_{\rm x}$=10$^{41}$-10$^{45}$ erg s$^{-1}$ \citep{vasudevan2007}.  
When X-ray emission is not available, we obtain bolometric luminosities from the  [OIII] line assuming  the  conversion factor of 3500 given in \citet{heckman2004}.
 This value is obtained for sources non-corrected by dust extinction. 
 Assuming an efficiency $\epsilon$ of 10\%  \citep{marconi2004},  the black hole accretion rate  is given by 
\begin{eqnarray}
{\rm BHAR}[{\rm M}_{\odot} {\rm yr}^{-1}]=0.15~ \frac{L_{\rm bol}} {10^{45}},
\end{eqnarray}
where the bolometric luminosity  $L_{\rm bol}$ is in erg s$^{-1}$ \citep{alexander2012}. 
For mergers, we find bolometric luminosities in the range 
2.4$\times$10$^{43}$--7.4$\times$10$^{46}$ erg s$^{-1}$ and BHARs of  3.6$\times$10$^{-3}$--11.1 M$_{\odot}$ yr$^{-1}$ (median of 0.07 M$_{\odot}$ yr$^{-1}$). For non-mergers, bolometric luminosities are 
2.2$\times$10$^{42}$--1.3$\times$10$^{48}$ erg s$^{-1}$ and BHARs are in the range 3.4$\times$10$^{-4}$--194 M$_{\odot}$ yr$^{-1}$ (median of 0.04 M$_{\odot}$ yr$^{-1}$). 
For star-forming galaxies, we also measure the BHAR in starbursts and main sequence galaxies (Table \ref{tbl_bhar}) to analyze if starbursting galaxies have higher black hole accretion rates. We find that starbursting merging galaxies  have higher accretion into their black holes compared to non-mergers  (starbursts and main sequence)  and main sequence merging galaxies, although precaution has to be taken with this result  due to the poor statistics of starbursts.
 Figure \ref{fig_main_seq_bhar} presents the position of merging and non-merging galaxies in the main sequence diagram  of star-forming galaxies.  Galaxies are color-coded by the intensity of the  BHARs.  The diagram shows that  the accretion rate into black holes  increases  with redshift.

%%%%%%%.  FIGURE   plot_log-bhar-sfr_z.py 
\begin{figure*}[!htbp]
\begin{center}
\includegraphics[angle=0,scale=0.55]{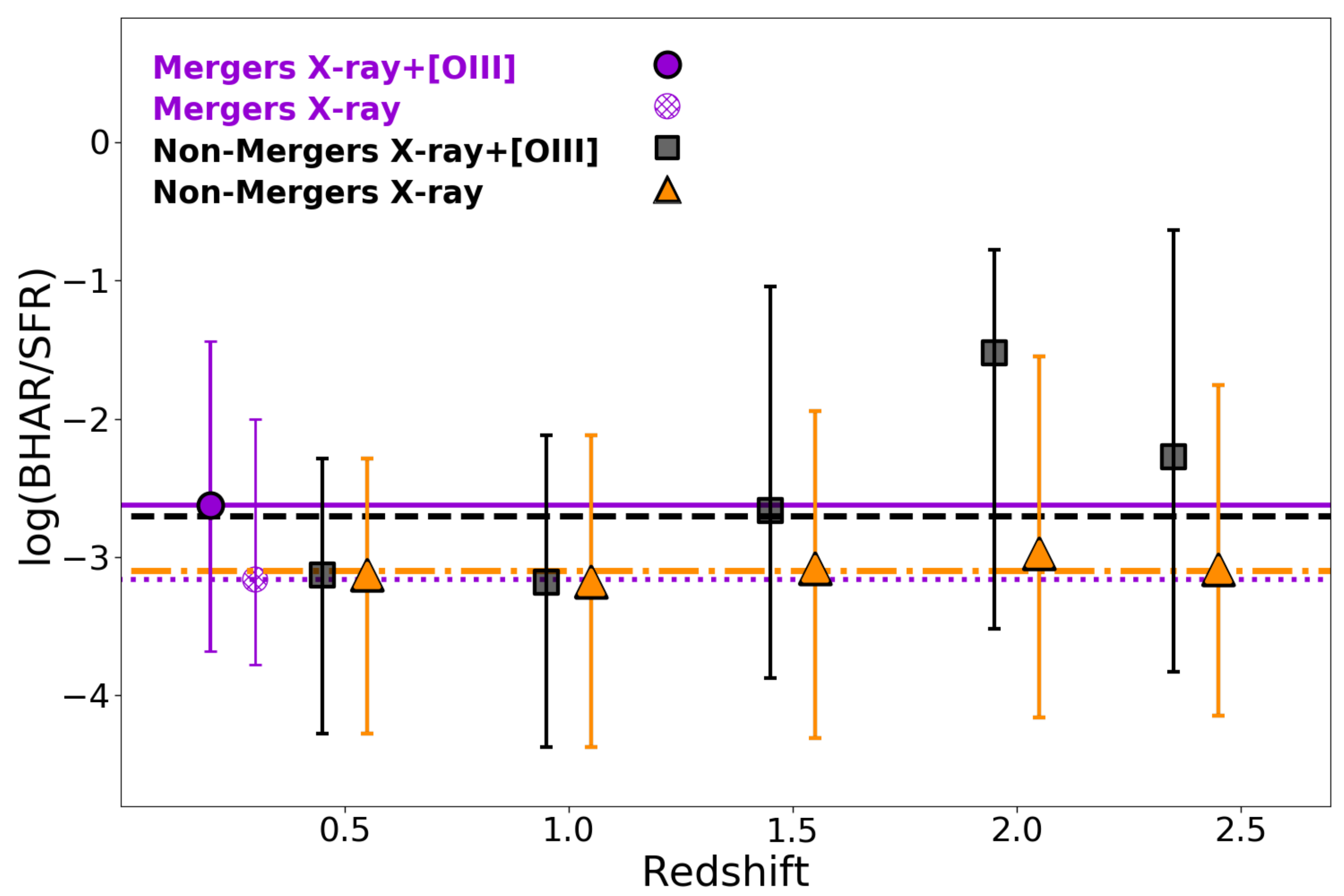}
\caption{  Ratio of the black hole accretion rate (BHAR) over the star formation rate (SFR) as a function of redshift for galaxies with X-ray or optically-selected AGNs.
The solid line and the filled purple circle  correspond to the median of the ratio over the whole redshift range  for BHARs obtained from X-ray and/or [OIII] luminosities, while  the dashed filled circle and the dotted purple line is the median for BHARs obtained using only the X-ray luminosities.  For mergers,  we assume that the merging galaxy with the highest star formation rate is the X-ray-selected AGN.
Squares and triangles are the median of the ratios in different redshift bins for non-mergers.  
   The  black dashed and the orange dotted-dashed  lines indicate  the median of the BHAR/SFR ratio for non-merging galaxies over the whole redshift range. 
   Lower and upper error bars correspond to the 15$^{\rm th}$ and 85$^{\rm th}$ percentiles of the distributions, respectively. 
\label{fig_ratio_bhar-sfr}}
\end{center}
\end{figure*}

%%%%%%%%%%.  FIGURE plot_bhar_various.py  
\begin{figure*}[!htbp]
\begin{center}
\includegraphics[angle=0,scale=0.59]{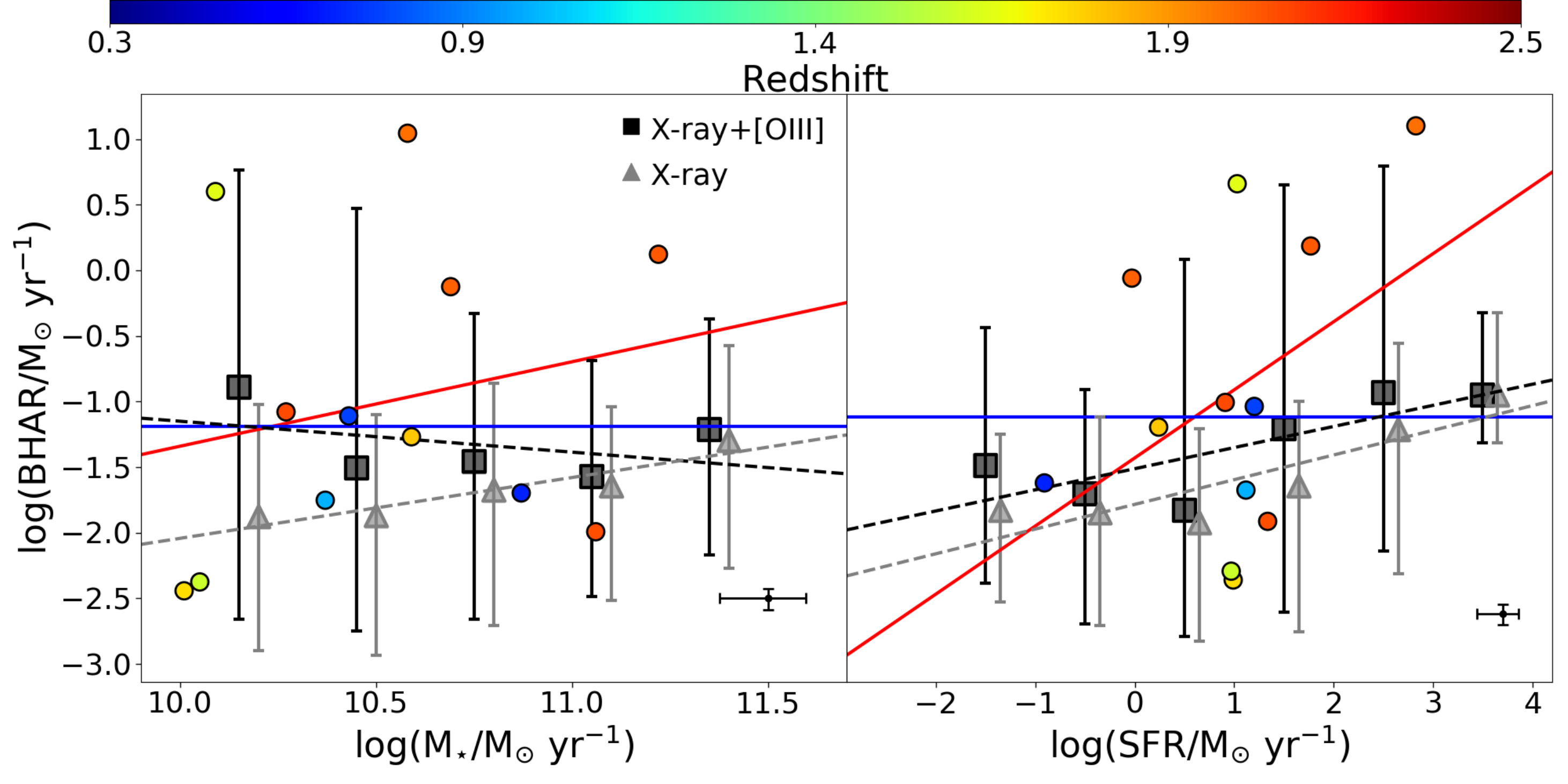}
\caption{ Black hole accretion rate as a function of stellar mass (left) and star formation rate (right) over the whole redshift range. Circles show the individual position of the 12 merging galaxies with X-ray and/or [OIII] emission.  Black squares and gray triangles show the median of the BHAR in different bins of stellar mass/star formation rate for non-merging galaxies 
with BHARs obtained from the   X-ray and/or [OIII]  emission (BHAR$_{\rm XR+[OIII]}$) and X-ray emission only (BHAR$_{\rm XR}$), respectively. 
 Lower and upper error bars correspond to the 15$^{\rm th}$  and 85$^{\rm th}$  percentiles of the distributions, respectively.  The solid red line is a linear fit of the position in this plot for mergers, while the dashed black and gray lines are linear fits of the median values for non-merging galaxies.
 The solid blue lines correspond to the median of the log(BHAR)=-1.19  for mergers assuming no correlation at all between the BHAR and the stellar mass  and the star formation rate.   Median of the uncertainties are displayed at the bottom right in each panel.
\label{fig_relations_bhar_mstar_sfr}}
\end{center}
\end{figure*}

 We investigate the evolution of the relative growth rate of a galaxy and its central black hole at different cosmic times in mergers and non-mergers by measuring the ratio  BHAR/SFR  as a function of redshift. 
 Figure  \ref{fig_ratio_bhar-sfr}  presents the median values of this ratio  for mergers and non-mergers in the redshift range 0.3$<$$z$$<$2.5.   For non-mergers, we show the median of the ratio in different redshift bins and find an increment with redshift, although not significant given the high dispersion (black squares in Fig.  \ref{fig_ratio_bhar-sfr}).  The median of the BHAR/SFR ratios in the whole redshift range for non-mergers is 1.9$\times 10^{-3}$  (dashed black line in Fig. \ref{fig_ratio_bhar-sfr}). 
 Since the BHARs obtained from [OIII] luminosities are uncorrected by dust attenuation, we also analyze the ratio by using only the galaxies with  BHARs obtained  from X-ray emission. 
  The median of the ratio in this case show no evolution with redshift (orange triangles in Fig. \ref{fig_ratio_bhar-sfr}), similar to the lack of evolution shown  in \citet{silverman2009} for a sample of $z$$\lesssim$1 AGNs.
   However, we find a total median value for the ratio in non-mergers of  8.0$\times$10$^{-4}$  (dashed-dotted orange line in Fig. \ref{fig_ratio_bhar-sfr}) while \citet{silverman2009} find 1.9$\times$10$^{-2}$. 
   This difference might be possible due to the different star-formation rate indicators used in the measurements. \citet{silverman2009} use the [O II] line, which is known to be a tracer of current star-formation activity, while  in this work, we use star formation rates obtained from UV+IR, which trace long-term star formation.
 Since in 10 merging systems the X-ray AGNs are not resolved, we estimate the BHAR/SFR making two assumptions: the merging galaxy with the lowest and then with the highest star formation rate in a system  is the AGN source. Differences in the measured values are listed in Table \ref{tbl_bhar}. 
When assuming that the merging galaxy with the highest SFR is the AGN source, we find a similar median on the BHAR/SFR ratio than in non-mergers  (2.4$\times 10^{-3}$ and 1.9$\times 10^{-3}$ for mergers and non-mergers, respectively). These values are similar to  the values of the local relation M$_{\rm BH}$/M$_{\rm bulge}$=1.5$\times 10^{-3}$ and the value predicted from the  simulations of  \citet{silk2013}. 

 Figure \ref{fig_relations_bhar_mstar_sfr} presents the relation between the BHAR with the star-formation rate and stellar mass for mergers and non-mergers. 
   We perform linear fits of the type $\log({\rm BHAR})=\alpha \log(x) + \beta$ where x is log(M$_{\star}$) or log(SFR).
The slopes in the BHAR vs stellar mass diagram  are $\alpha$=0.38, -0.23, and 0.45 for mergers, non-mergers with 
BHARs obtained from the   X-ray and/or [OIII]  emission (BHAR$_{\rm XR+[OIII]}$), and non-mergers with BHARs from 
X-ray emission only (BHAR$_{\rm XR}$), respectively. 
The slopes in the BHAR vs star formation rate diagram are $\alpha$=0.4, 0.17, 0.19 for mergers, non-mergers with BHAR$_{\rm XR+[OIII]}$, and non-mergers with BHAR$_{\rm XR}$, respectively. 
  We find no correlation between the BHAR and the SFR and the stellar mass in non-mergers.  In mergers, it seems that a correlation exists, but precaution has to be taken with this result  due to the poor statistics and the unresolved emission of the X-ray AGNs.

\begin{deluxetable*}{lcccccc}
\tabletypesize{\footnotesize}
\tablecaption{Values of the black hole accretion rate (BHAR) and its relation with the star formation rate (SFR) for optical or X-ray-selected AGNs. In parenthesis are the number of galaxies taken for the measurements.  \label{tbl_bhar}}
\tablewidth{0pt}
\tablehead{
\colhead{Median Values} &  \multicolumn{2}{c}{Mergers} & \multicolumn{1}{c}{non-mergers} \\
\colhead{} &   \colhead{Lower SFR$^{a}$ (N$_{\rm gal}$) } & \colhead{Higher SFR$^{b}$ (N$_{\rm gal}$)}  &  \colhead{ (N$_{\rm gal}$)}   
}
\startdata
BHAR  [M$_{\odot}$ yr$^{-1}$]                        &  0.07 (12) & 0.07 (12)   &  0.06 (596)          \\
BHAR Starbursts   [M$_{\odot}$ yr$^{-1}$]          &  11.1 (1) & 5.6 (2)    &  0.02 (64)        \\
BHAR Main Sequence  [M$_{\odot}$ yr$^{-1}$]  &  0.02 (7) & 0.08 (9)    &  0.06 (417)          \\
BHAR/SFR                                                          &  1.3$\times 10^{-2}$ (12)  & 2.4$\times 10^{-3}$ (12)    &  1.9$\times 10^{-3}$ (596)         \\   
BHAR/SFR (X-ray only)                                       &   7.6$\times 10^{-4}$  &    7.0$\times 10^{-4}$            &       8.0$\times 10^{-4}$ (447)  \\
BHAR/SFR Starbursts                                  &  1.6$\times 10^{-2}$  & 8.4$\times 10^{-3}$  (2)  &   1.0$\times 10^{-4}$   (64)  \\  
BHAR/SFR   MS.                                          &  1.3$\times 10^{-3}$ & 9.3$\times 10^{-4}$  (9)& 2.2$\times 10^{-3}$  (417)\\
 \enddata
\tablenotetext{a}{Assuming that the merging galaxy in a system with the lowest star formation rate is the AGN source}
\tablenotetext{b}{Assuming that the merging galaxy in a system  with the highest star formation rate is the AGN source}
 \end{deluxetable*}

%%%%%    DISCUSSION  %%%%%  
\section{Discussion} \label{sec_discussion}

Numerical simulations suggest that during a merger, there is a flow of large quantities of gas  from kpc scales toward the central regions of the galaxies. When a merger is at coalescence,  the system experiences  the fastest black hole growth and the highest star formation activity \citep[e.g.][]{springel2005, hopkins2008, debuhr2011}.  
This is in agreement with observations mostly of  $z$$\leq$1 galaxies, where AGNs, especially the brightest ones, have been found in advanced merger stages \citep[e.g.][]{cotini2013, goulding2018, ellison2019}. There are  findings that show an excess of AGNs in galaxy pairs compared to isolated galaxies  \citep[e.g.][]{silverman2011, satyapal2014, ellison2019}.  
Although our sample of 0.3$<$$z$$<$2.5 mergers contains galaxies with projected nuclei separations between 3$-$15 kpc, i.e., close to  the coalescence stage, we find no such AGN excess  (total AGN fractions of  16.4$\pm ^{5.0} _{3.1}$\% and 15.4$\pm 0.6$\%, for mergers and non-mergers respectively) and no clear increment in the AGN fraction with decreasing projected separation between merging galaxies (Fig. \ref{fig_frac_z_sep}).
 In addition, in \citet{silva2018}, using the same sample of mergers used in this study, we found no difference between the star formation activity between mergers compared to non-mergers.  
%
%  the main point is the competition between gravitational torques due to internal dynamics and due to the galaxy interaction
%
%
The mass assembly of galaxies seems to  occur in two main modes: merging and 
cold accretion through intergalactic filaments  \citep{kauffmann1993, keres2005, dekel2009, bournaud2009, lhuillier2012}.
In the merging scenario, cold gas in the galaxies loses angular momentum, increases it turbulence, and flows inwards the nuclei activating rapid star formation and accretion into the SMBHs \citep[e.g.][]{hernquist1989, barnes1996}.
%\citep[e.g.][]{spri
In the cold gas accretion picture, gas streams feed galaxies. This accretion triggers instabilities and turbulence in the gas inducing mass inflows which allow the formation of star-forming clumps and the growth of the central SMBH \citep{bournaud2011, bournaud2012}. 
At high redshift the universe was denser and had higher gas fractions than today \citep{tacconi2010, daddi2010, scoville2014, tacconi2018}. Therefore, the active parts of high redshift galaxies  are denser, more gas rich, and more turbulent than local galaxies. 
 It has been measured that  gas fraction in massive spiral galaxies increases from $\sim$10\% at $z$$\sim$0 to $\sim$50\% at $z$$\sim$2 and so gravitational instabilities and velocity dispersions are also higher at high redshift  ($\sigma$=30-100 km s$^{-1}$) compared to low redshift spirals \citep[$\sigma \sim$10 km s$^{-1}$;][]{andersen2006, epinat2010, swinbank2011, stott2016}. 
As cold streams continuously feed galaxies, gas fractions decrease with cosmic time,  resulting
 in a declining in the star formation and black hole activity since $z$$\sim$2  \citep{silverman2008, ebrero2009, rodighiero2010, aird2010, vandevoort2011, cucciati2012}.
The lack of an excess of star formation and AGN activity as a result of merging compared to the overall population of high redshift galaxies  found in \citet{silva2018} and in this work, is possible due to the inefficiency of high redshift mergers \citep{kaviraj2013, lofthouse2017, fensch2017}. 
Since high redshift galaxies are already highly turbulent and unstable due to the cosmic cold gas accretion, it is difficult for a merger to further increase their turbulence.
Therefore the increment due to merging in the star formation and AGN activity at high redshift  might be mild compared to the increment that mergers produce in less turbulent, low redshift galaxies.
On the other hand, we find that AGNs from mergers follow the same evolution as those resulting from non-merging galaxies, with an AGN fraction increasing with redshift. This is  in agreement with 
what has been  pointed out by e.g. \citet{ellison2019} and \citet{draper2012}, which suggest that mergers are  probably not the dominant mechanism that trigger AGN activity,  since the fraction of mergers  with AGN hosts has been mostly found to be low \citep[e.g.][]{villforth2017, marian2019}.

We find a higher AGN fraction in star-forming than in quiescent galaxies for both mergers and non-mergers (Fig. \ref{fig_frac_agn}, middle) in agreement with the increment of AGN activity in mergers containing at least one star-forming galaxy, i.e. wet and mix mergers (Fig. \ref{fig_frac_mergtype}). In addition,  starbursts galaxies present a   higher AGN fraction for both merging and non-merging galaxies compared to the total AGN fraction.  
Star-forming galaxies are sources containing higher cold gas fractions than quiescent galaxies\footnote{We do not have measurement of cold gas content in this sample of mergers.}  and high redshift starbursts are possibly the result of galaxies containing high levels  of cold gas  \citep{scoville2016, oteo2017}. Therefore, it is an expected consequence that these populations will have higher AGN activity, since the accretion into a super massive black hole  and star formation are processes that need a cold gas supply to take place.  
Our results are in agreement with the findings of  \citet{santini2012}, which  report evidence of higher average star formation activity  in X-ray-selected AGN hosts compared to a control sample of inactive galaxies, including both quiescent and star-forming galaxies at 0.5$<$$z$$<$2.5.

Although we find  similar BHARs between merging and non-merging galaxies, we find that merging starbursting galaxies have higher BHARs  and BHAR/SFR ratios compared to starbursts in non-mergers and main sequence galaxies in both mergers and non-mergers. This result suggests that the merging of galaxies with high cold gas content (i.e. starbursts)  are even more efficient at triggering AGNs.  
Similar to our results,  \citet{rodighiero2019} find higher BHARs in starbursts, with a factor of 3 enhancement  compared to normal star-forming galaxies for a sample of 1.5$<$$z$$<$2.5 galaxies with a great diversity of star-forming properties.  
 In their Figure 3, they present  predictions of an  enhancement of  BHAR and BHAR/SFR during the
merger phase based on the hydrodynamical models of \citet{dimatteo2005} and \citet{hopkins2012}, which is in agreement with our findings (see Table \ref{tbl_bhar}).  

Hydrodynamical simulations presented in \citet{volonteri2015b} suggest that nuclear SFR ($<$100 pc) have higher correlation with BHAR  than galaxy-wide SFR, since the central regions are more influenced by the SMBH.  The minimum physical size of the apertures for which the stellar mass and star-formation rate were measured in our sample  is $\sim$1.5 kpc (aperture of 0\farcs3 at $z$=0.3). Therefore, we are analyzing the large scale SFR  where no clear correlation between the BHAR and SFR is expected. 
 On the other hand, the majority of our galaxies for which the BHAR was measured, contain  low-luminosity AGNs  with 67\% and 68\% of the merging and non-merging galaxies having L$_{bol}$$<$10$^{45}$ erg s$^{-1}$, respectively. For galaxies with $L_{bol}$ obtained  from X-ray emission only the fractions  increase to 80\% and 90\% for mergers and non-mergers, respectively. 
 Therefore, the lack of a direct correlation between BHAR and SFR and stellar mass in our sample  is in agreement with previous studies toward low-luminosity AGNs   \citep{shao2010, lutz2010, rovilos2012, rosario2012}, where the star-formation and the black hole are dominated by secular processes.  
 In our sample of mergers,  it is difficult to analyze whether there is a correlation or not due to poor statistic. However, at first glance, it seems that mergers have higher correlation of BHAR  with SFR than non-mergers, as a possible result of the influence of the merging process (Fig. \ref{fig_relations_bhar_mstar_sfr}, \ref{fig_app1}, and \ref{fig_app2}). In the coalescence phase, the merger should have its peak  in star formation and AGN activity and  a correlation should emerge.  
  Although we find no direct correlation between the BHAR and SFR, the mean of the ratios of BHAR/SFR is almost constant with redshift and with the same order of magnitude that the local M$_{\rm BH}$/M$_{\rm bulge}$ relation. 
  This result suggests that the mean growth time of the SMBHs and the mean growth time of the stellar content in the galaxies  are related in the same way even at high redshifts.

  When we assume that the source of unresolved X-ray or radio-selected AGNs are the two merging galaxies in a system, we find a higher AGN fraction in mergers compared to non-mergers (29.2$\pm ^{6.5}_{4.5}$\% vs 15.4$\pm$0.6\%). Although it is possible that some of these unresolved AGNs are dual, the observed fraction of dual AGNs with respect to the total AGN population has been found to be low.
  For instance in the nearby universe the fraction of dual AGNs has been found to be lower than 10\% \citep{liu2011,koss2012,imanishi2020}.   Numerical simulations suggest that the fraction of dual AGNs increases with redshift \citep{volonteri2016, rosasguevara2019}. However, recently \citealt{silverman2020} find a fraction of dual AGNs of 0.26\% with no evidence of evolution up to z$<$4.5. 
  The low observed fraction of dual AGNs is possibly because the AGN phase itself is ephemeral and its ignition depends on the properties of the host galaxy, such as stellar mass and gas fractions.  Therefore, the chance of finding  two active SMBHs is low. Since the number of confirmed dual AGNs is still scarce, it is difficult to determine the true frequency of dual AGNs and therefore estimate how many of our non-resolved AGNs are potentially dual.
   Higher resolution observations are needed to probe whether each galaxy hosts a unique AGN.   Some of the future projects that will help in the identifications of resolved AGNs, especially  in samples of merging galaxies at high redshifts, are the following (see \citealt{derosa2020} for detailed descriptions):   {\it Lynx} \footnote{https://wwwastro.msfc.nasa.gov/lynx/} and {\it AXIS} \citep{mushotzky2018}  are missions under consideration by NASA to provide  sensitive (at least 10 times better than {\it Chandra}) and high  resolution ($\sim$0\farcs5) observations of X-ray AGNs.  The James Webb Space Telescope (JWST)  will provide both high angular resolution imaging and spectroscopy at mid-infrared wavelengths. Specifically,  the Infrared Field Unit Near InfraRed Spectrograph (NIRSpec), which operates over a wavelength range of 0.6 to 5.3 $\mu$m, will have a resolution $\sim$5$\times$ better than the highest resolution spectrograph in {\it Spitzer}. Finally,  the Square Kilometre Array (SKA) will have resolutions of 0\farcs5-1\arcsec  in the frequency range 0.95-1.76 GHz, which is $\geq$3 times better than current JVLA 1.4 GHz observations.

%%%%%   SUMMARY AND CONCLUSIONS   %%%%%  
\section{Summary and Conclusions} \label{sec_conclusion}

We quantify the presence of AGNs  in a sample of merging systems at 0.3$<$$z$$<$2.5 with projected separations of 3-15 kpc between the galaxy nuclei. We analyze the sample obtained in the first paper of the series \citet{silva2018}, which was obtained from the CANDELS/3D-HST catalogs using the peak-finding algorithm developed in \citet{lackner2014}. We identify  AGN activity in mergers and non-mergers (galaxies not selected as mergers) by measuring the X-ray and radio emission, their mid-Infrared colors, and their [OIII]+H$\beta$ optical line emission.  
Our findings can be summarized as follows:\\
%
%\begin{itemize}

$\bullet$  Among galaxies with adequate measurements to find potential AGNs, we find a similar AGN fraction between  mergers (16.4$\pm^{5.0}_{3.1}$\%) and non-mergers (15.4$\pm$0.6\%). The fraction in merging galaxies is obtained by assuming that only one of the merging galaxies is the AGN source (there are 16 merging systems with at least one AGN galaxy) when the AGN is obtained from unresolved observations.  
This result is in agreement with the lack of difference in the star-formation activity between mergers and non-mergers  found in \citet{silva2018}, indicating that mergers are inefficient in increasing the star-formation and AGN activity at high redshifts.  Due to the higher availability of cold gas at earlier cosmic times,
the potential excess of star formation and nuclear activity as a result of merging fade with respect to the overall population of high redshift galaxies.\\

$\bullet$  We find a higher AGN fraction in star-forming   than in quiescent galaxies in both mergers and 
non-mergers.  In addition, we find a higher AGN fraction in wet and mix mergers compared to dry mergers. These results indicate that mergers with at least one star-forming galaxy and star-forming non-mergers galaxies have a higher 
AGN incidence.  
Starburst galaxies, especially  those in mergers, are the most efficient at triggering AGN activity due to the high AGN fraction and BHARs they present compared to the rest of the population, even other star-forming galaxies.  \\

$\bullet$  We find no clear  correlation between the black hole accretion rate and the star-formation rate and stellar mass in mergers and non-mergers. 
Mergers seem to have a higher level of correlation with star formation than non-mergers, which might indicate the merging process is influencing the star formation and AGN activity.  Note that we do not probe the coalescence phase, for which a correlation is expected on theoretical grounds.\\

$\bullet$  Higher resolution observations are needed to resolve the emission of  X-ray and radio-selected AGNs to disentangle if our sample contain dual AGNs. \\

%\end{itemize}

We thank the anonymous referee for useful comments which  helped to  improve this paper.

\vspace{0.5cm}

DM acknowledges support from the Japan Society for the Promotion of Science (JSPS) through the JSPS Invitational Fellowships for Research in Japan (Short-term, ID S19135), and from the Tufts University Faculty Research Fund (FRF).

%%%%%%%%%%%%%%%%%%%%%%%%%%%
%\clearpage
\begin{appendix}

\section{Relation Black hole accretion rate with stellar mass and star formation rate}

\begin{figure}[!htbp]
\begin{center}
\includegraphics[angle=0,scale=0.50]{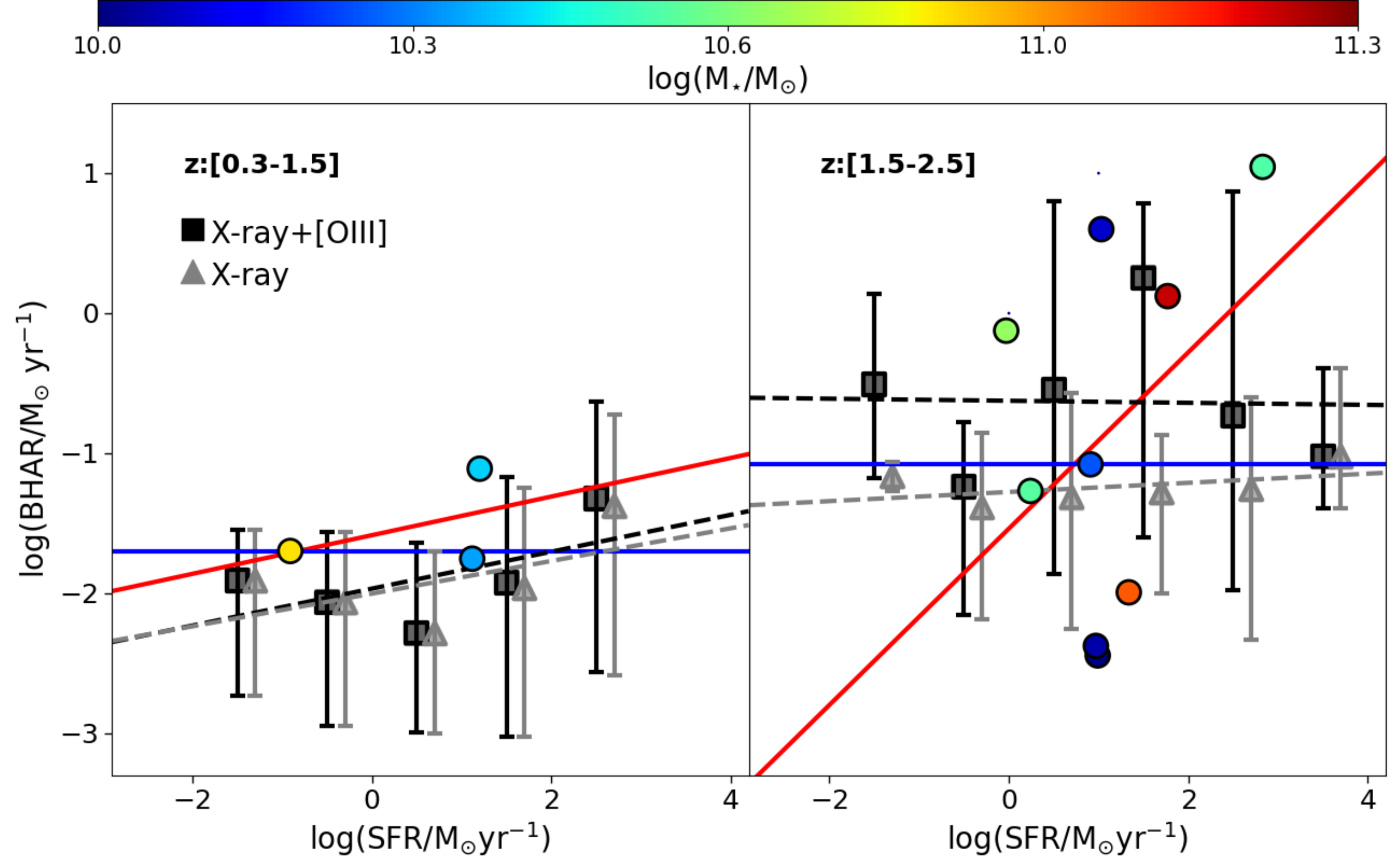}
\caption{Black hole accretion rate as a function of star formation rate  (color coded by stellar mass) in two redshift bins. Circles show the individual position of the 12 merging galaxies with X-ray and/or [OIII] emission.  Black squares and gray triangles show the median of the BHAR in different bins of star formation rate for non-merging galaxies with BHARs obtained from the X-ray and/or [OIII] emission  (BHAR$_{\rm XR+[OIII]}$)  and X-ray emission only (BHAR$_{\rm XR}$), respectively. 
Lower and upper error bars correspond to the 15$^{\rm th}$ and 85$^{\rm th}$ percentiles of the distributions, respectively.  The solid red line is a linear fit of the position in this plot for mergers, while the dashed black and gray lines are linear fits  of the median values for non-merging galaxies. 
In the  redshift range 0.3$\leq$$z$$<$1.5, the slopes of the linear fits are  $\alpha$=0.01, 0.13, and  0.11 for mergers, non-mergers with BHAR$_{\rm XR+[OIII]}$, and non-mergers with BHAR$_{\rm XR}$, respectively.
 In the redshift range 1.5$\leq$$z$$<$2.5, the slopes of the linear fits are  $\alpha$=-0.015, -0.008, and 0.03 for mergers, non-mergers with BHAR$_{\rm XR+[OIII]}$, and non-mergers with BHAR$_{\rm XR}$, respectively. The solid blue line correspond to the median of the log(BHAR) for mergers assuming no correlation at all between the BHAR and the SFR.  These median values are -1.70 and -1.08 for the lower and the higher redshift range, respectively.
\label{fig_app1}}
\end{center}
\end{figure}

\begin{figure}[!htbp]
\begin{center}
\includegraphics[angle=0,scale=0.50]{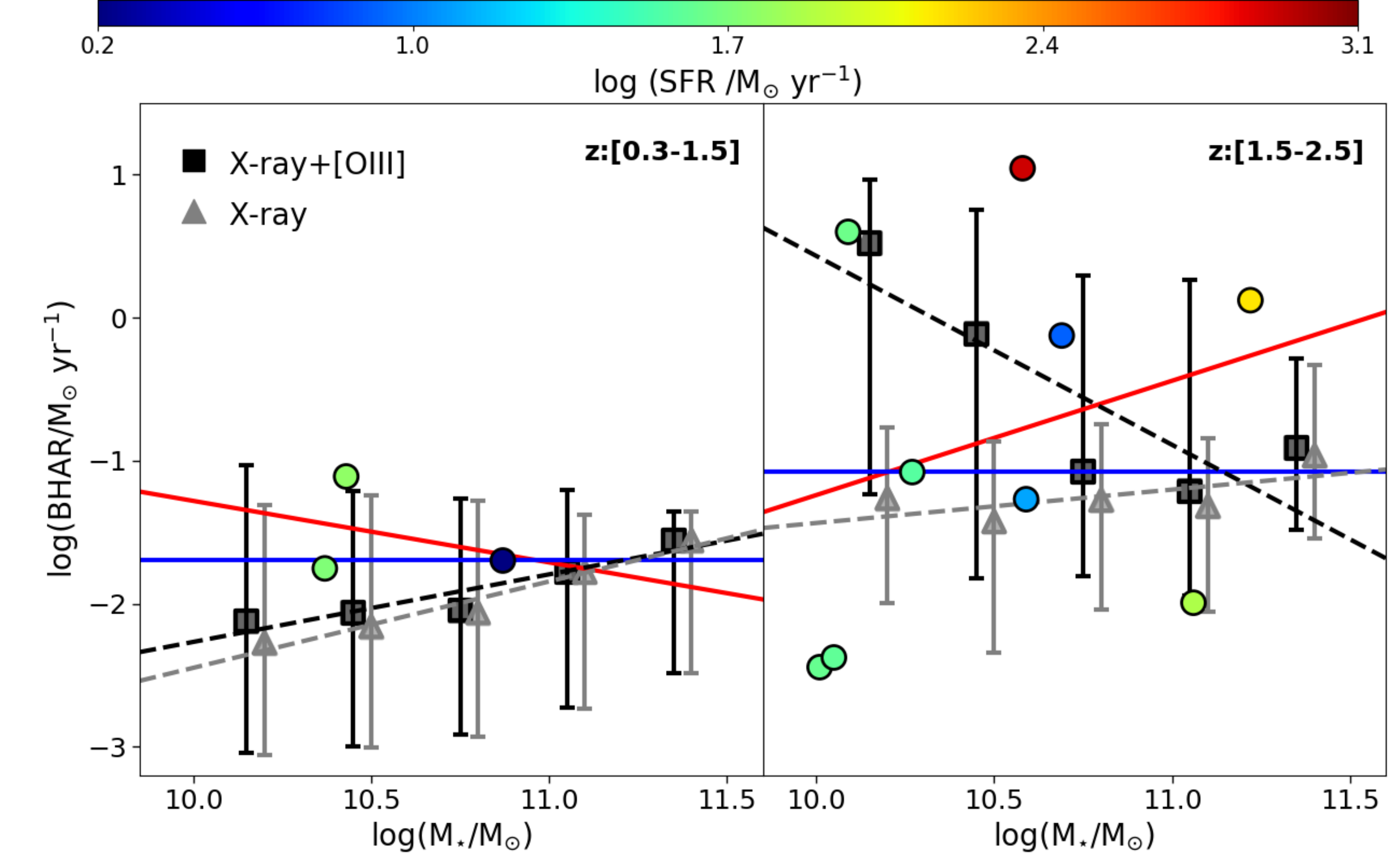}
\caption{Black hole accretion rate as a function of stellar mass (color coded by the star formation rate)  in two redshift bins. Circles show the individual position of the 12 merging galaxies with X-ray and/or [OIII] emission. Black squares and gray triangles show the median of the BHAR in different bins of stellar mass  for non-merging galaxies with BHARs obtained from the   X-ray and/or [OIII]  emission (BHAR$_{\rm XR+[OIII]}$) and X-ray emission only (BHAR$_{\rm XR}$), respectively. 
Lower and upper error bars correspond to the 15$^{\rm th}$ and 85$^{\rm th}$ percentiles of the distributions, respectively.  The solid red line is a linear fit  of the position in this plot for mergers, while the dashed black and gray lines are linear fits of the median values for non-merging galaxies.
In the  redshift range 0.3$\leq$$z$$<$1.5, the slopes of the linear fits are  $\alpha$=0.19, 0.47, and 0.6 for mergers, non-mergers with BHAR$_{\rm XR+[OIII]}$, and non-mergers with BHAR$_{\rm XR}$, respectively.
In the redshift range 1.5$\leq$$z$$<$2.5,  the slopes of the linear fits are  $\alpha$=-0.02, -1.3, and 0.23 for mergers, non-mergers with BHAR$_{\rm XR+[OIII]}$, and non-mergers with BHAR$_{\rm XR}$, respectively. The solid blue lines correspond to the median of the log(BHAR) for mergers assuming no correlation at all between the BHAR and the SFR.  These median values are -1.70 and -1.08 for the lower and the higher redshift range, respectively.
\label{fig_app2}}
\end{center}
\end{figure}

In this Appendix, we  present plots that relate the black hole accretion rate in galaxies with the star formation rate and the stellar mass.  We only plot merging and non-merging galaxies with BHARs obtained from the X-ray and/or the  [OIII] emission. 
For X-ray AGNs selected in mergers (i.e. the AGN is unresolved), we assume that the merging galaxies with the higher star formation rate is the AGN source.
We also repeat our results by using BHARs obtained using only the X-ray emission of the AGNs. 
In Figures  \ref{fig_app1} and   \ref{fig_app2} we perform linear fits of the type $\log({\rm BHAR})=\alpha \log(x) + \beta$ where x is log(M$_{\star}$) or log(SFR).

\end{appendix}

\clearpage

%%%%%  BIBLIOGRAFY %%%%%

\end{document}